Article

# ASSESSING THE INFLUENCE OF CYBERSECURITY THREATS AND RISKS ON THE ADOPTION AND GROWTH OF DIGITAL BANKING: A SYSTEMATIC LITERATURE REVIEW

**Md. Waliullah[1]; Md Zahin Hossain George[2]; Md Tarek Hasan3; Md Khorshed Alam[4]; Mosa Sumaiya Khatun Munira[5];Noor Alam Siddiqui[6];**

[1]Master of Science in Management Information Systems, College of Business, Lamar University, Texas, USA
Email: mwaliullah@lamar.edu

[2]MSc in Cybersecurity, Department of Professional Security Studies, New Jersey City University, USA
Email: mgeorge1@njcu.edu

[3]MS in Cybersecurity, Department of Professional Security Studies, New Jersey City University, USA
Email: mhasan2@njcu.edu

[4]MS in Cybersecurity, Department of Professional Security Studies, New Jersey City University, USA
Email: malam@njcu.edu

[5]MBA, Scott College of Business, Indiana State University, USA
Email: skmunira@gmail.com

[6]Master of Science in Management Information Systems, College of Business, Lamar University, USA
E-mail: noor.siddiqui440@gmail.com

## ABSTRACT

*The rapid digitalization of banking services has significantly transformed financial transactions, offering enhanced convenience and efficiency for consumers. However, the increasing reliance on digital banking has also exposed financial institutions and users to a wide range of cybersecurity threats, including phishing, malware, ransomware, data breaches, and unauthorized access. This study systematically examines the influence of cybersecurity threats on digital banking security, adoption, and regulatory compliance by conducting a comprehensive review of 78 peer-reviewed articles published between 2015 and 2024. Using the Preferred Reporting Items for Systematic Reviews and Meta-Analyses (PRISMA) methodology, this research critically evaluates the most prevalent cyber threats targeting digital banking platforms, the effectiveness of modern security measures, and the role of regulatory frameworks in mitigating financial cybersecurity risks. The findings reveal that phishing and malware attacks remain the most commonly exploited cyber threats, leading to significant financial losses and consumer distrust. Multi-factor authentication (MFA) and biometric security have been widely adopted to combat unauthorized access, while AI-driven fraud detection and blockchain technology offer promising solutions for securing financial transactions. However, the integration of third-party FinTech solutions introduces additional security risks, necessitating stringent regulatory oversight and cybersecurity protocols. The study also highlights that compliance with global cybersecurity regulations, such as GDPR, PSD2, and GLBA, enhances digital banking security by enforcing strict authentication measures, encryption protocols, and real-time fraud monitoring. Despite these advancements, financial institutions face ongoing challenges in balancing security, usability, and regulatory compliance, which impact consumer trust and digital banking adoption. The review underscores the need for a multi-layered security strategy that integrates encryption, AI-driven fraud prevention, blockchain security, and robust regulatory frameworks to ensure the long-term resilience of digital banking. This study contributes to the growing body of knowledge on financial cybersecurity, offering insights into emerging threats, risk mitigation strategies, and policy recommendations for securing digital financial ecosystems.*

## KEYWORDS

*Digital Banking; Cybersecurity Threats; Risk Mitigation; Consumer Trust; Financial Technology (FinTech)*

## INTRODUCTION

The financial services sector has experienced significant transformation with the rapid adoption of digital banking, which has fundamentally changed how banking transactions are conducted (Akhtar & Das, 2019). Digital banking encompasses a wide range of services, including online banking, mobile banking applications, and digital payment systems that enable users to access financial services conveniently and securely (Chandra sekhar & Kumar, 2023). The increasing reliance on digital banking has been driven by technological advancements such as artificial intelligence (AI), blockchain, and cloud computing, which offer enhanced efficiency and user-friendly interfaces (Elia et al., 2022). However, as digital banking continues to grow, concerns related to cybersecurity threats have become more pronounced, posing challenges to financial institutions and consumers alike (Ahmad et al., 2024). Cybersecurity threats, including phishing, malware, identity theft, and data breaches, have the potential to compromise sensitive financial information, leading to substantial financial losses and reputational damage (Cele & Kwenda, 2024). The increasing frequency and sophistication of cyberattacks require robust security frameworks to ensure the integrity, confidentiality, and availability of banking services. Moreover, cybersecurity threats in digital banking have been widely documented in academic literature, with studies highlighting their far-reaching implications on financial security, regulatory compliance, and consumer trust (Castelli et al., 2016). Among the most common cyber threats, phishing attacks remain a critical concern, where fraudsters use deceptive emails or fake websites to trick users into revealing their banking credentials (Chandra sekhar & Kumar, 2023). A study by (Boon-itt, 2015) found that phishing remains one of the leading causes of unauthorized financial transactions, as attackers exploit human vulnerabilities rather than technological loopholes. In addition to phishing, malware and ransomware attacks pose significant risks by infiltrating banking systems and encrypting critical data, demanding a ransom for decryption (Ahmad et al., 2024). According to Chauhan et al. (2022), malware-based cyber threats often target mobile banking applications, exploiting weak security mechanisms to gain unauthorized access to users' financial information. These persistent threats highlight the urgent need for financial institutions to implement proactive cybersecurity measures to prevent fraudulent activities and ensure uninterrupted banking operations.

**Figure 1: Overview of Digital Transformation in Banking Industry**

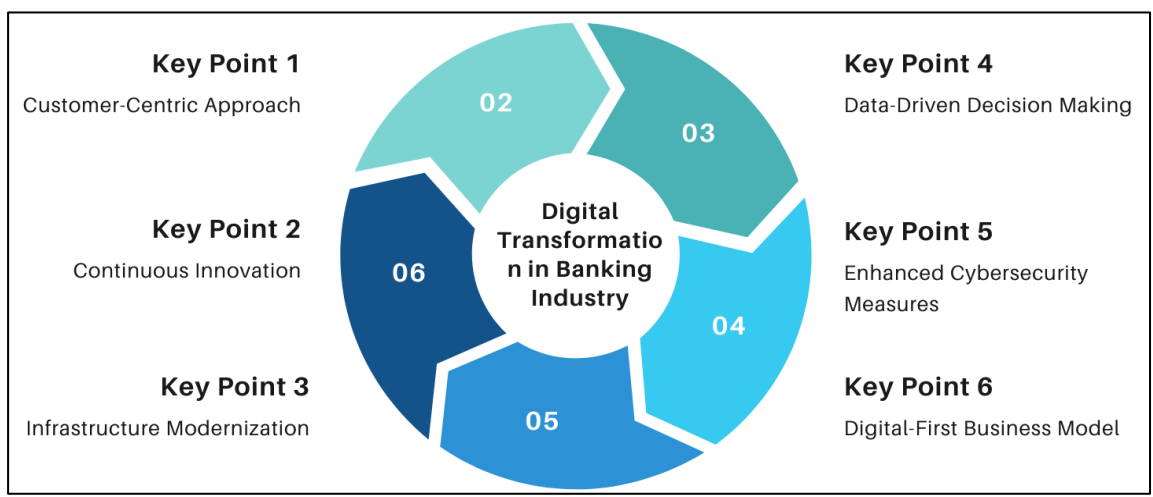

The role of consumer trust in digital banking adoption has been extensively studied, demonstrating that cybersecurity concerns directly impact user behavior and banking preferences (Elia et al., 2022). Security breaches can erode consumer confidence, discouraging individuals from using digital banking platforms due to fear of financial fraud (Bapat, 2017). Mbama and Ezepue (2018) found that perceived cybersecurity risks negatively affect the adoption of mobile banking services, as users prioritize security over convenience. Furthermore, a study by Larsson and Viitaoja (2017) revealed that users with prior negative experiences related to cyber fraud exhibit higher levels of reluctance in embracing digital banking solutions. This suggests that financial institutions must not only focus on implementing security measures but also communicate their security policies effectively to reassure customers. Transparency in cybersecurity practices, such as clear policies on fraud protection, multi-factor authentication (MFA), and encryption protocols, plays a crucial role in shaping consumer trust and encouraging digital banking adoption (Bapat, 2017). To counter cybersecurity threats, financial institutions have adopted various risk mitigation strategies to enhance the security of digital banking platforms. The implementation of multi-factor authentication, biometric security, and end-to-end encryption has become a standard approach in securing online transactions (Chauhan et al., 2021). A study by Ahmad et al. (2024) found that the integration of biometric authentication methods, such as fingerprint and facial recognition, significantly improves the security of digital banking applications by reducing reliance on passwords that are susceptible to hacking. Additionally, Elia et al. (2022) emphasized the importance of AI-driven fraud detection systems that analyze transaction patterns in real time to identify and prevent suspicious activities. Blockchain technology has also emerged as a promising solution for enhancing the security of financial transactions, offering decentralized and tamper-resistant mechanisms to prevent data breaches and fraud (Boon-itt, 2015). As financial institutions continue to integrate these technological advancements, the effectiveness of cybersecurity measures will play a critical role in shaping the future of digital banking security.

**Figure 2: Cybersecurity in Digital Banking: Trust, Technologies, and Regulations**

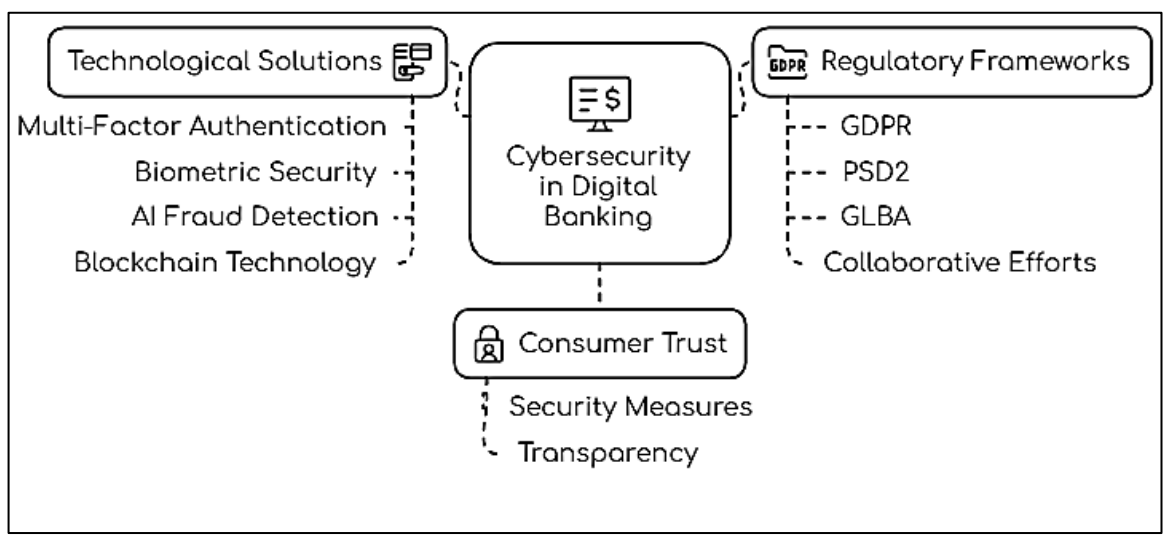

Regulatory frameworks and compliance measures are essential in ensuring the security and resilience of digital banking systems. Governments and financial regulatory bodies worldwide have introduced stringent cybersecurity policies aimed at protecting consumer data and preventing cybercrimes (Calderaro & Craig, 2020). For example, the European Union's General Data Protection Regulation (GDPR) and the Payment Services Directive 2 (PSD2) impose strict guidelines on data protection and secure authentication processes to safeguard online financial transactions (Pan & Fan, 2021). Similarly, the United States' Gramm-Leach-Bliley Act (GLBA) mandates financial institutions to implement robust cybersecurity measures to protect sensitive consumer information (Wright et al., 2009). Aldasoro et al. (2020) found that compliance with cybersecurity regulations not only improves the security posture of financial institutions but also enhances consumer trust and confidence in digital banking services. However, financial institutions must remain vigilant in ensuring that regulatory requirements are effectively implemented and updated to counter emerging cyber threats. Moreover, the global financial landscape is becoming increasingly interconnected, necessitating collaboration among financial institutions, technology providers, and regulatory bodies to strengthen cybersecurity frameworks (Brechbühl et al., 2010). Cybercriminals are continuously evolving their attack strategies, employing advanced persistent threats (APTs) and social engineering techniques to exploit vulnerabilities in digital banking systems (Marqués et al., 2021). A review by Sullivan and Burger (2017) highlighted the growing sophistication of cyberattacks, emphasizing the need for a multi-layered cybersecurity approach that integrates technical defenses, employee training, and consumer awareness initiatives. Financial institutions must adopt a proactive stance by investing in cybersecurity research and development to stay ahead of emerging threats (Teece, 2018). Moreover, cross-industry collaborations, information-sharing networks, and public-private partnerships play a crucial role in developing collective security strategies that enhance the resilience of digital banking ecosystems. The objective of this study is to systematically examine the influence of cybersecurity threats and risks on the adoption and growth of digital banking by synthesizing existing literature on the subject. Specifically, this review aims to identify the key cybersecurity threats affecting digital banking, such as phishing, malware, identity theft, and data breaches, and assess their impact on consumer trust and banking security. Additionally, the study seeks to evaluate the effectiveness of various risk mitigation strategies, including multi-factor authentication, blockchain technology, and AI-driven fraud detection systems, in enhancing digital banking security. Another objective is to analyze the role of regulatory frameworks and compliance measures, such as GDPR, PSD2, and GLBA, in addressing cybersecurity risks and protecting consumer financial data. Furthermore, this study aims to explore how cybersecurity concerns influence consumer perceptions and adoption behavior in digital banking. By achieving these objectives, this review provides a comprehensive understanding of the challenges and strategies associated with securing digital banking platforms, contributing to the ongoing discourse on financial cybersecurity and risk management.

**LITERATURE REVIEW**

The rise of digital banking has led to significant transformations in the financial sector, offering greater convenience, accessibility, and efficiency for both consumers and financial institutions. However, these advancements have also introduced new cybersecurity risks that threaten the security and trustworthiness of digital financial transactions (Cherdantseva et al., 2016). The literature on digital banking security has extensively examined various cyber threats, including phishing, malware attacks, identity theft, and data breaches, which have posed challenges for financial institutions and consumers alike (Gomathi & Jayasri, 2022). As digital banking services continue to evolve, it is imperative to understand how these cybersecurity risks influence consumer adoption, institutional resilience, and regulatory responses (Emara & Zhang, 2021). This section synthesizes existing literature on the intersection of cybersecurity and digital banking, highlighting key security threats, risk mitigation strategies, regulatory frameworks, and their impact on consumer trust and adoption. This structured review provides a detailed examination of the current state of digital banking security, identifying gaps in existing research while offering insights into emerging security strategies and regulatory approaches.

**Transition from traditional banking to digital banking**

The transition from traditional banking to digital banking has fundamentally transformed the financial services industry, reshaping the way individuals and businesses conduct financial transactions (Wang et al., 2020). Traditional banking, characterized by in-person visits to physical branches and reliance on paper-based transactions, has increasingly been replaced by digital alternatives, driven by advancements in financial technology (FinTech) and changing consumer preferences (Marqués et al., 2021). This transformation has been facilitated by the proliferation of online and mobile banking platforms, enabling customers to access banking services remotely and conduct transactions with greater convenience (Al-Shari & Lokhande, 2023). Studies have highlighted that the adoption of digital banking is fueled by several factors, including improved internet accessibility, the expansion of mobile technology, and the emergence of cloud computing-based financial services (Al-Shari & Lokhande, 2023; Senyo et al., 2022; Warjiyono et al., 2019). Despite the advantages associated with digital banking, scholars have raised concerns regarding cybersecurity threats, as financial institutions and consumers are increasingly exposed to sophisticated cyberattacks, including phishing, ransomware, and identity theft (Chang et al., 2020; Mehrotra & Menon, 2021; Senyo et al., 2022). The shift from traditional banking to digital platforms necessitates robust cybersecurity measures to protect sensitive financial data and maintain consumer trust (He et al., 2020).

The rapid growth of online and mobile banking services has been a key driver of financial inclusion and accessibility, allowing individuals from diverse demographic and geographic backgrounds to engage with banking services in ways that were previously unavailable (Singh et al., 2020; Wahab et al., 2021). Online banking platforms enable customers to perform a wide range of transactions, including fund transfers, bill payments, and loan applications, without the need for physical interactions (Doumpos et al., 2023). Mobile banking, in particular, has experienced exponential growth due to increased smartphone penetration, enhanced mobile internet infrastructure, and the rise of digital payment ecosystems (Doumpos et al., 2023; Warjiyono et al., 2019). However, studies indicate that while online and mobile banking services enhance user convenience and operational efficiency, they also present significant security challenges, including malware infections, fraudulent transactions, and unauthorized access to personal accounts (Elia et al., 2022). Researchers emphasize that financial institutions must implement stringent cybersecurity protocols, such as multi-factor authentication, end-to-end encryption, and AI-driven fraud detection systems, to mitigate security risks and sustain consumer confidence (Doumpos et al., 2023).

**Figure 3: Traditional vs Digital Banking**

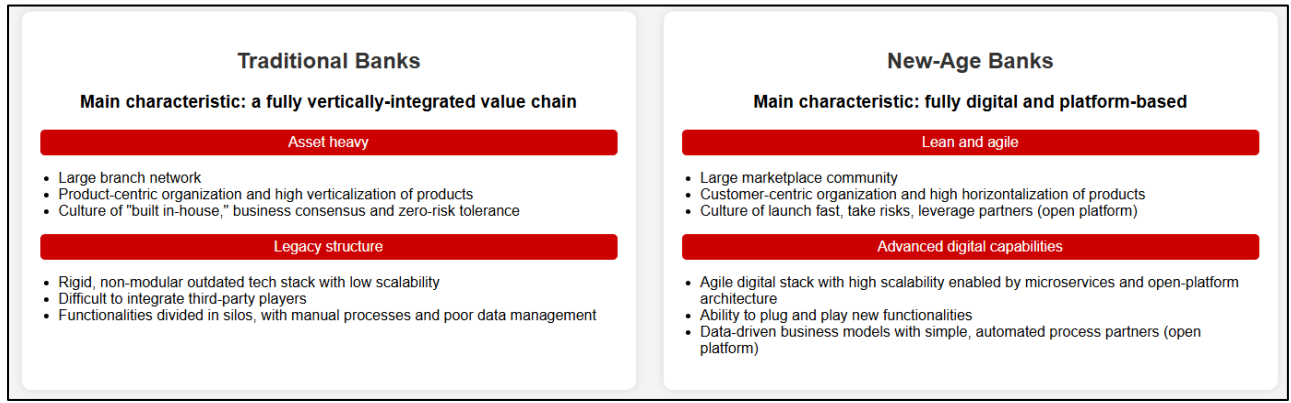

Despite its numerous benefits, digital banking is increasingly vulnerable to security threats due to the complexity and interconnectivity of modern financial networks. Cybersecurity vulnerabilities associated with digital banking platforms stem from various attack vectors, including weak authentication protocols, software vulnerabilities, and inadequate risk management strategies (Khan et al., 2023). Studies reveal that cybercriminals exploit security loopholes in online banking systems to execute financial fraud, steal sensitive customer information, and disrupt banking operations through denial-of-service (DoS) attacks (Calderaro & Craig, 2020; Khan et al., 2023). A study by Campbell (2019) highlighted that data breaches in financial institutions not only result in financial losses but also damage institutional reputation and consumer trust. Researchers emphasize that financial institutions must continuously update their security frameworks to address evolving cyber threats, implementing robust authentication mechanisms, AI-based anomaly detection, and blockchain-driven security models (Calderaro & Craig, 2020; Duran & Griffin, 2020). Furthermore, compliance with regulatory frameworks, such as the General Data Protection Regulation (GDPR), the Payment Services Directive 2 (PSD2), and the Gramm-Leach-Bliley Act (GLBA), plays a crucial role in enhancing cybersecurity resilience and safeguarding customer data in digital banking (Shabbir et al., 2022; Taylor et al., 2020).

The increased dependency on digital transactions has further amplified security risks, as financial transactions are now primarily conducted through online banking portals, mobile applications, and digital payment systems (Ring, 2014). Studies indicate that the growing reliance on digital banking has led to an increase in cyber threats, as attackers target financial systems to exploit security weaknesses and gain unauthorized access to user credentials (Paul & Wang, 2019; Warjiyono et al., 2019). Researchers argue that financial institutions must implement comprehensive cybersecurity risk management strategies, incorporating biometric authentication, secure communication protocols, and AI-driven fraud prevention tools, to mitigate security vulnerabilities (Ring, 2014). Additionally, consumer awareness and cybersecurity literacy play a critical role in reducing the risks associated with digital banking, as informed users are less likely to fall victim to phishing scams, identity theft, and fraudulent transactions (Shabbir et al., 2022; Taylor et al., 2020). Given the complexities of modern cybersecurity threats, financial institutions must adopt proactive security policies and continuously monitor emerging threats to ensure the stability and security of digital banking services (Khan et al., 2022; Paul & Wang, 2019).

**Phishing Attacks and Social Engineering**
Phishing attacks and social engineering tactics remain among the most prevalent cybersecurity threats in the digital banking sector, exploiting human vulnerabilities rather than technological loopholes (Tn & Shailendra Kulkarni, 2022). Phishing is a form of cyber fraud in which attackers use deceptive emails, messages, or fake websites to trick users into disclosing sensitive information such as login credentials, personal identification numbers (PINs), or financial details (Arshad et al., 2021). Social engineering tactics leverage psychological manipulation to exploit users' trust and convince them to take actions that compromise their security, such as clicking malicious links or downloading infected attachments (Ahmad et al., 2024; Patil et al., 2022). A study by Mehbodniya et al. (2021) found that phishing attacks have evolved beyond email-based schemes, incorporating voice phishing (vishing), SMS phishing (smishing), and spear phishing techniques to specifically target banking customers and employees. The effectiveness of phishing attacks lies in their ability to bypass traditional security measures, making it crucial for financial institutions to implement strong security awareness training and robust authentication mechanisms (Cele & Kwenda, 2024; Johri & Kumar, 2023).

**Figure 4: Phishing Attack Mindmap**

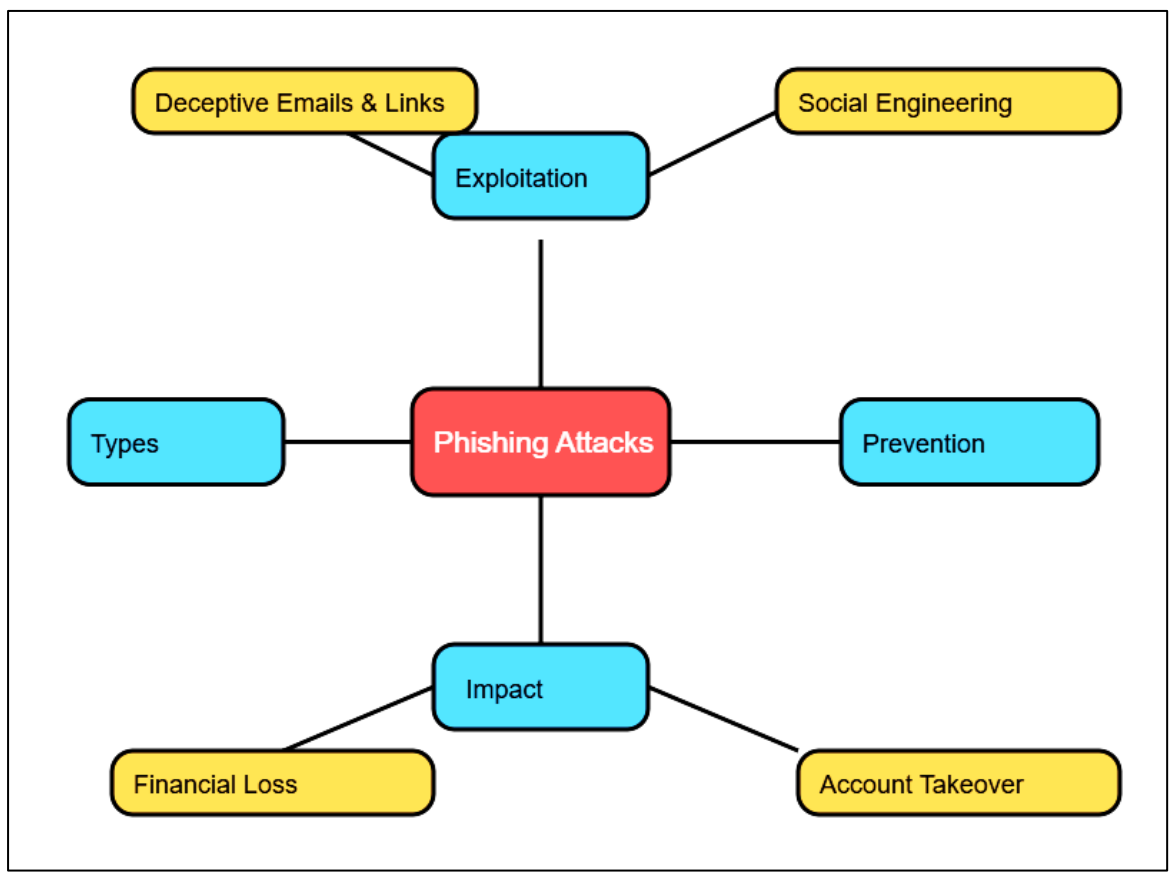

The exploitation of human vulnerabilities through deceptive emails and fraudulent websites has significantly contributed to the rise of financial fraud in online banking (Ali, 2019). Attackers design phishing websites that closely resemble legitimate banking portals, often using fake domain names and security certificates to deceive unsuspecting users (Al-Khater et al., 2020; Cele & Kwenda, 2024). According to a study by Mehbodniya et al. (2021), attackers exploit cognitive biases and urgency tactics, such as fake security alerts or account suspension warnings, to pressure users into providing sensitive credentials. (Nawa et al., 2021) noted that phishing campaigns often rely on large-scale email distribution networks, targeting thousands of users simultaneously in hopes of compromising a fraction of them. Financial institutions have attempted to mitigate phishing risks by incorporating advanced email filtering mechanisms, anti-phishing software, and user education programs (Arshad et al., 2021; Patil et al., 2022). However, despite these measures, phishing remains a persistent challenge due to the constant evolution of attack techniques and the adaptability of cybercriminals (Cele & Kwenda, 2024).

Phishing remains a significant contributor to online banking fraud, leading to substantial financial losses for individuals and institutions (Tn & Shailendra Kulkarni, 2022). A report by Arshad et al. (2021) found that phishing-related fraud accounts for a major portion of cybercrime in the financial sector, with attackers stealing credentials to gain unauthorized access to digital banking accounts. Once attackers obtain banking credentials, they engage in fraudulent transactions, unauthorized withdrawals, or identity theft, severely impacting victims (Patil et al., 2022). Some phishing schemes involve account takeovers, where cybercriminals change login details and lock users out of their own accounts, making recovery difficult (Mehbodniya et al., 2021). Advanced persistent phishing attacks leverage artificial intelligence and deepfake technology to impersonate legitimate banking representatives, increasing the success rate of fraudulent schemes (Johri & Kumar, 2023; Mehbodniya et al., 2021). Given the financial and reputational risks associated with phishing-related fraud, banking institutions must continually invest in sophisticated fraud detection mechanisms, behavioral biometrics, and AI-driven threat intelligence systems to identify and prevent phishing attacks before they cause significant damage (Ali, 2019; Cele & Kwenda, 2024). Several studies have explored different strategies for preventing phishing attacks in online banking, emphasizing the importance of a multi-layered security approach (Al-Khater et al., 2020; Aljeaid et al., 2020; Nawa et al., 2021). Implementing multi-factor authentication (MFA), such as biometric verification and one-time passwords (OTP), has proven to be effective in preventing unauthorized access even if phishing attackers obtain user credentials (Tn & Shailendra Kulkarni, 2022). Additionally, AI-powered fraud detection systems that analyze user behavior patterns and flag login anomalies have been increasingly adopted by financial institutions to detect phishing attempts in real time (Arshad et al., 2021). Patil et al., (2022) emphasized that consumer education and awareness programs play a crucial role in reducing phishing success rates, as informed users are less likely to fall for deceptive schemes. Mehbodniya et al. (2021) further noted that collaboration between financial institutions, cybersecurity firms, and regulatory bodies is essential in strengthening anti-phishing defenses through data-sharing agreements and standardized security protocols. While no single measure can fully eliminate phishing threats, a combination of technical safeguards, continuous monitoring, and user awareness initiatives remains the most effective strategy in mitigating the risks associated with phishing in online banking (Aljeaid et al., 2020; Nawa et al., 2021).

**Malware and Ransomware Attacks**

Malware and ransomware attacks have emerged as major cybersecurity threats to digital banking, targeting financial institutions and individual users through sophisticated infiltration techniques (Mehbodniya et al., 2021). Malware refers to malicious software designed to infiltrate banking systems, steal sensitive data, and execute unauthorized financial transactions (Johri & Kumar, 2023). Cybercriminals deploy various types of malware, including banking Trojans, keyloggers, and remote access tools (RATs), to gain control over banking networks and user accounts (Cele & Kwenda, 2024). A study by Ali (2019) revealed that financial institutions are particularly vulnerable to malware attacks due to their reliance on interconnected digital infrastructures, which provide multiple entry points for attackers. Malware often infiltrates banking systems through phishing emails, malicious attachments, or compromised mobile applications, exploiting software vulnerabilities and weak security protocols (Javaid, 2013). Research indicates that financial institutions must adopt robust endpoint security measures, regular software updates, and advanced threat detection technologies to prevent malware infiltration and minimize financial losses (Humayun et al., 2020; Kimani et al., 2019; McGraw, 2013).

**Figure 5: Ransomware Attack Growth by Quarter Across the World**

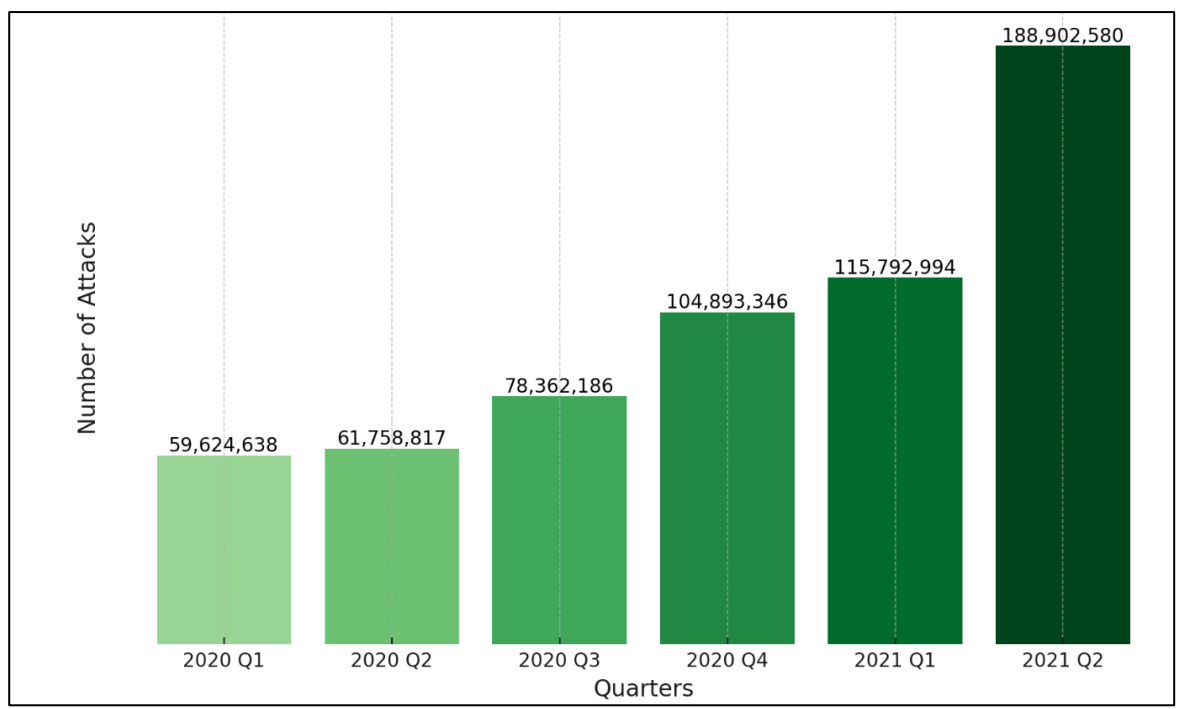

Ransomware attacks on financial institutions have intensified in recent years, causing severe disruptions and financial losses by encrypting critical data and demanding ransom payments for decryption (Bouveret, 2019; Kopp et al., 2017; Muhammad Mohiul et al., 2022). Ransomware infiltrates banking systems through malicious software that locks users out of their networks and encrypts essential files, rendering financial transactions and customer access impossible (Akintoye et al., 2022; Bouveret, 2018b). Case studies of major ransomware attacks illustrate the devastating impact on digital banking operations, with some financial institutions experiencing prolonged service outages, reputational damage, and substantial financial losses (Bhuiyan et al., 2024; Kopp et al., 2017; Oladapo et al., 2021). For instance, a ransomware attack on a major global bank in 2020 resulted in the compromise of millions of customer records and led to a regulatory investigation into the bank's cybersecurity preparedness (Aklima et al., 2022; Bouveret, 2019; Shahan et al., 2023). Another high-profile case in 2021 saw attackers using sophisticated ransomware variants, such as Ryuk and Conti, to extort financial institutions by threatening to release stolen data on the dark web (Makeri, 2017). These incidents underscore the need for digital banking institutions to strengthen their cybersecurity defenses, implement real-time ransomware detection mechanisms, and develop comprehensive incident response plans (Liu et al., 2022).

Detecting malware threats in digital banking requires the deployment of advanced cybersecurity solutions that leverage artificial intelligence (AI), machine learning, and behavioral analytics to identify anomalies and prevent unauthorized access (Bouveret, 2018a; Kopp et al., 2017). AI-driven cybersecurity tools have proven effective in analyzing vast amounts of banking transaction data to detect unusual patterns indicative of malware infections (Oladapo et al., 2021). Additionally, threat intelligence platforms enable financial institutions to share real-time cyber threat information, allowing them to stay ahead of emerging malware variants (Ali, 2019; Bouveret, 2018a). Studies have also emphasized the role of endpoint security solutions, such as antivirus software, firewalls, and intrusion detection systems (IDS), in mitigating malware risks (Bouveret, 2018b, 2019). According to Kopp et al. (2017), financial institutions that employ AI-based threat detection and real-time network monitoring experience a significant reduction in malware-related security breaches. However, researchers also warn that cybercriminals are continuously evolving their tactics, necessitating an adaptive approach to cybersecurity that integrates both technological defenses and human vigilance (Akintoye et al., 2022). Moreover, Preventing malware and ransomware attacks in digital banking requires a multi-layered security strategy that encompasses strong authentication mechanisms, regular security audits, and employee awareness training (Gomes et al., 2022; Kopp et al., 2017). Implementing multi-factor authentication (MFA), biometric verification, and blockchain-based security protocols enhances the resilience of banking networks against malware intrusions (Bouveret, 2018b). Additionally, financial institutions must conduct frequent security audits and penetration testing to identify vulnerabilities before cybercriminals exploit them (Humayun et al., 2020; Liu et al., 2022). Employee training programs play a crucial role in minimizing human error, as research has shown that unintentional actions, such as clicking on malicious links or downloading infected attachments, are among the primary entry points for malware infections (Makeri, 2017). Collaboration between banks, cybersecurity firms, and regulatory bodies is also essential in establishing industry-wide best practices for malware prevention and ensuring compliance with stringent cybersecurity regulations (Kimani et al., 2019). As the threat landscape continues to evolve, maintaining a proactive and adaptable cybersecurity posture remains crucial for safeguarding digital banking systems against malware and ransomware attacks (Akintoye et al., 2022).

**Identity Theft and Account Takeover Fraud**

Identity theft and account takeover fraud pose significant risks to digital banking security, as cybercriminals employ increasingly sophisticated methods to steal banking credentials and gain unauthorized access to financial accounts (Sharma & Tandekar, 2018). One of the most prevalent methods involves phishing attacks, where attackers use deceptive emails, fraudulent websites, or text messages to trick users into revealing their login credentials (Gomes et al., 2022; Hossain et al., 2024; Sharma & Tandekar, 2018). Keylogging malware and credential-stealing Trojans are also widely used by cybercriminals to capture sensitive user information, often without the victim's knowledge (Ali, 2019; Merhi et al., 2019). Additionally, cybercriminals exploit vulnerabilities in public Wi-Fi networks, where unsuspecting users connect to unsecured hotspots, exposing their banking credentials to man-in-the-middle attacks (Jim et al., 2024; Kesswani & Kumar, 2015; Kopp et al., 2017). A study by Bouteraa et al., (2022) found that social engineering tactics, such as impersonating bank officials or using deepfake technology to bypass authentication processes, have significantly increased in frequency. The widespread availability of stolen banking credentials on the dark web further exacerbates the issue, allowing cybercriminals to purchase and use compromised accounts for fraudulent activities (Younus et al., 2024; Riad & Elhoseny, 2022). These attack vectors demonstrate the need for financial institutions to continuously strengthen security protocols to prevent unauthorized access and identity fraud.

Identity theft has severe consequences for consumer trust in digital banking, leading to financial losses, reputational damage, and reluctance to use online banking services (Karthik, 2024; Siddiki et al., 2024). Research indicates that once consumers experience identity theft, they often exhibit higher levels of distrust toward digital banking platforms, preferring traditional banking methods or withdrawing from online banking altogether (Adedoyin Tolulope et al., 2024; Karthik, 2024). Studies have shown that consumers who have fallen victim to identity theft often face challenges in recovering lost funds, as fraud investigations and reimbursement processes can be lengthy and complicated (Berkman et al., 2018; Vagle, 2020). A study by Bernik (2014) found that identity theft victims are more likely to switch financial institutions, negatively impacting customer retention and brand loyalty. Furthermore, widespread reports of account takeover fraud undermine the reputation of banks, increasing regulatory scrutiny and operational costs associated with fraud mitigation (Liu et al., 2022; McGraw, 2013). Research highlights that financial institutions must adopt proactive measures to reassure customers about the security of digital banking systems, such as offering fraud protection policies, implementing real-time fraud alerts, and improving customer service response times (Makeri, 2017).

**Figure 6: Storyboard: Identity Theft & Account Takeover Fraud**

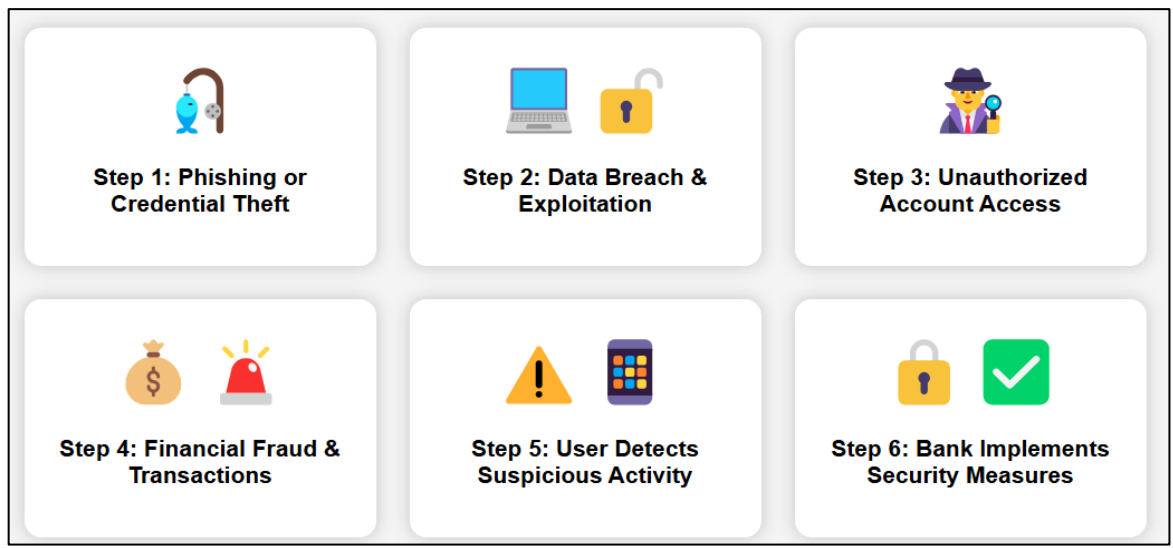

Technological advancements in identity verification and fraud prevention have played a critical role in mitigating identity theft and account takeover fraud in digital banking (Choo, 2011). Multi-factor authentication (MFA), which requires users to verify their identities through additional security layers such as biometrics, one-time passwords (OTP), and behavioral analytics, has proven effective in reducing unauthorized access (Bernik, 2014). A study by Vagle (2020) found that financial institutions that incorporate biometric authentication methods, such as facial recognition and fingerprint scanning, experience significantly lower rates of account takeover fraud. Additionally, AI-driven fraud detection systems analyze user behavior patterns, transaction anomalies, and location-based activity to detect fraudulent activities in real time (Berkman et al., 2018). Blockchain technology has also emerged as a promising solution for securing identity verification processes, offering decentralized authentication mechanisms that prevent credential tampering and unauthorized modifications (Kox, 2013). Research indicates that integrating these technological solutions with secure communication protocols, such as end-to-end encryption and tokenization, further enhances digital banking security (Karthik, 2024). Despite advancements in identity verification technology, financial institutions must continuously evolve their security frameworks to address emerging threats related to identity theft and account takeover fraud (Olukunle Oladipupo et al., 2024). Studies suggest that banks should implement risk-based authentication, which dynamically adjusts security measures based on user behavior and transaction risk levels (Adedoyin Tolulope et al., 2024). AI-driven fraud detection models can analyze transaction history, device usage, and geolocation data to flag suspicious activities before fraudulent transactions occur (Bechara & Schuch, 2020). Research also highlights the importance of real-time fraud monitoring systems that provide immediate alerts to users when unusual transactions or login attempts are detected (Choo, 2011). Furthermore, collaboration between financial institutions, regulatory bodies, and cybersecurity firms is essential in developing industry-wide standards for identity verification and fraud prevention (Olukunle Oladipupo et al., 2024). By adopting a multi-layered security approach that integrates cutting-edge technology, financial institutions can mitigate identity theft risks and enhance consumer trust in digital banking systems (Ali, 2019).

**Data Breaches and Unauthorized Access**

Data breaches and unauthorized access remain critical challenges in digital banking, exposing sensitive customer information and leading to significant financial and reputational losses for financial institutions (Kopp et al., 2017; Sharma & Tandekar, 2018). Cybercriminals employ sophisticated attack techniques such as SQL injection, credential stuffing, and insider threats to exploit vulnerabilities in banking networks and gain unauthorized access to confidential data (Akintoye et al., 2022; Kimani et al., 2019). A study by Ali (2019) found that financial institutions are frequent targets of data breaches due to the vast amounts of personal and financial information stored in their databases. Researchers also highlight that unauthorized access incidents often stem from weak authentication mechanisms, outdated software, and inadequate cybersecurity policies (Gomes et al., 2022; Humayun et al., 2020). According to Kimani et al. (2019), financial institutions that fail to implement multi-layered security frameworks are at a higher risk of data breaches, as cybercriminals continually evolve their attack strategies to bypass traditional security measures. The growing frequency of large-scale data breaches underscores the need for financial institutions to enhance their security protocols and adopt proactive threat mitigation strategies (McGraw, 2013). Case studies of major data breaches in financial institutions illustrate the severe consequences of cyber intrusions on the banking sector (Berkman et al., 2018). One of the most notable breaches occurred in 2019 when Capital One suffered a data breach that exposed the personal information of over 100 million customers, leading to regulatory fines and legal actions (Berkman et al., 2018; Vagle, 2020). Another high-profile case involved Equifax, where hackers exploited a software vulnerability in 2017, compromising the financial records of nearly 147 million individuals (Bechara & Schuch, 2020; Vagle, 2020). These breaches resulted in significant financial losses, legal penalties, and reputational damage, demonstrating the far-reaching impact of security lapses in digital banking (Malik & Islam, 2019; McGraw, 2013). Furthermore, a study by Liu et al. (2022)found that financial institutions that fail to adequately secure their data often face increased scrutiny from regulators, leading to compliance-related costs and operational disruptions. Case studies indicate that failure to implement timely security updates, encrypt sensitive data, and monitor network vulnerabilities contributes to large-scale data breaches in the financial sector (Bernik, 2014; Liu et al., 2022).

**Figure 7: Data Breaches and Unauthorized Access**

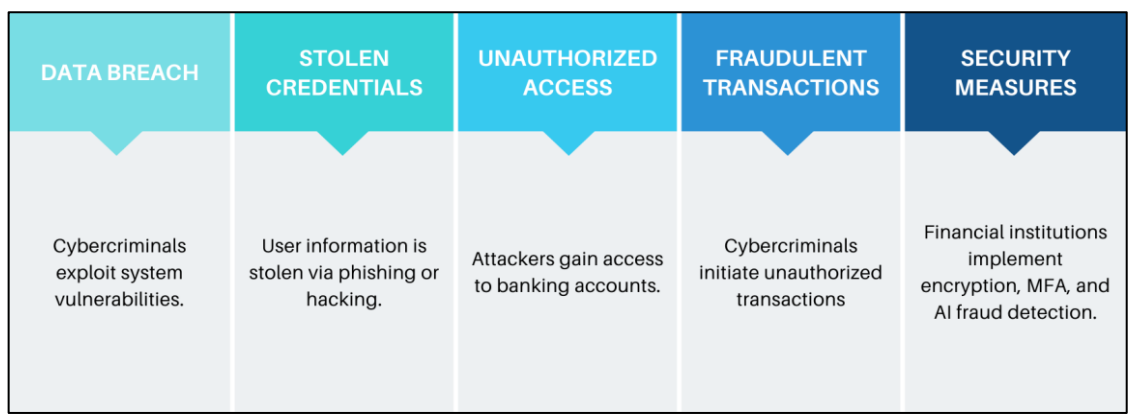

The consequences of data breaches extend beyond financial institutions, significantly affecting consumers who rely on digital banking services for financial transactions and account management (Choo, 2011). When personal information such as banking credentials, social security numbers, and credit card details are exposed, consumers become vulnerable to identity theft, fraudulent transactions, and financial losses (Javaid, 2013). A study by Karthik (2024) revealed that consumers who experience data breaches often lose trust in financial institutions, leading to reduced engagement with digital banking services and an increased preference for traditional banking methods. Additionally, compromised financial data can have long-term repercussions, including credit score damage and legal complications arising from fraudulent activities (Stewart & Jürjens, 2018; Wang et al., 2024). Research also indicates that financial institutions must take responsibility for ensuring consumer protection by offering fraud recovery programs, improving customer communication regarding security incidents, and enhancing user education on cybersecurity best practices (Adedoyin Tolulope et al., 2024). The reputational damage suffered by banks due to data breaches can lead to a decline in customer acquisition and retention rates, further exacerbating the financial impact of security incidents (Kox, 2013). To combat data breaches and unauthorized access, financial institutions employ various encryption and data protection techniques to safeguard sensitive information (Uddin et al., 2020). End-to-end encryption (E2EE) ensures that data transmitted between users and banking servers remains secure, preventing interception by cybercriminals (Kox, 2013; Vagle, 2020). A study by McGraw (2013) highlighted that tokenization, which replaces sensitive financial data with non-sensitive tokens, has become an effective approach in securing digital transactions. Additionally, multi-factor authentication (MFA) methods such as biometric verification, one-time passwords (OTP), and behavioral authentication enhance security by adding additional layers of protection against unauthorized access (Choo, 2011). Financial institutions also leverage blockchain technology to enhance data security, as decentralized ledgers prevent tampering and unauthorized modifications of financial records (Sharma & Tandekar, 2018). Studies emphasize that financial institutions must integrate advanced cybersecurity measures such as real-time threat detection, AI-driven fraud prevention, and regular security audits to mitigate the risks associated with data breaches and unauthorized access (Akintoye et al., 2022; Sharma & Tandekar, 2018).

**Cybersecurity Threats on Consumer Trust and Digital Banking Adoption**

Consumer trust in digital banking is significantly influenced by perceived cybersecurity risks, as concerns over financial fraud, identity theft, and data breaches deter users from fully embracing digital banking services (Ali, 2019). Research has shown that consumers evaluate the security of online banking platforms based on their perceived vulnerability to cyber threats, shaping their willingness to adopt digital financial services (Lee et al., 2020). A study by Kesswani and Kumar (2015) found that consumers with high-risk awareness are more likely to avoid online banking due to fear of financial loss or personal data exposure. Additionally, Choithani et al. (2022) highlighted that digital banking customers prioritize security over convenience, with many preferring traditional banking methods when cybersecurity concerns remain unaddressed. Studies indicate that financial institutions must implement transparent security measures, educate users on fraud prevention, and build customer confidence through robust authentication mechanisms to mitigate the negative impact of cybersecurity risks on digital banking adoption (Kopp et al., 2017).

The influence of prior security breaches on consumer behavior is a critical factor affecting digital banking adoption, as users who have experienced or heard about cyber fraud incidents exhibit heightened skepticism toward online financial transactions (Akintoye et al., 2022; Kopp et al., 2017). Research indicates that individuals who have fallen victim to cyberattacks often switch to alternative banking methods, such as in-person transactions or cash-based systems, to mitigate security risks (Javaid, 2013; Liu et al., 2022). According to Javaid (2013), widespread media coverage of major data breaches in financial institutions exacerbates consumer distrust, creating reluctance to engage with digital banking services. A study by Bechara and Schuch (2020) found that consumers with negative experiences related to unauthorized transactions or account takeovers tend to reduce their reliance on digital banking platforms, opting for financial institutions that offer enhanced security features. Financial organizations that fail to rebuild consumer confidence following security breaches face long-term reputational damage, decreased customer retention, and regulatory scrutiny (Bernik, 2014). Moreover, trust plays a central role in financial decision-making, particularly in the adoption of digital banking services, as users must feel assured that their personal and financial information is adequately protected (Uddin et al., 2020). Studies have emphasized that trust is built through institutional reputation, security transparency, and prior user experiences with digital banking systems (Akintoye et al., 2022). Liu et al. (2022) found that financial institutions that openly communicate their cybersecurity policies and fraud protection mechanisms tend to instill higher levels of trust among consumers. Sharma and Tandekar, (2018) argue that incorporating multi-layered authentication processes, encryption technologies, and AI-driven fraud detection systems enhances consumer confidence in digital banking security. Furthermore, research suggests that perceived trustworthiness significantly impacts consumer willingness to adopt emerging financial technologies such as mobile banking apps and blockchain-based transactions (McGraw, 2013; Sharma & Tandekar, 2018).

**Figure 8: Impact of Cybersecurity on Digital Banking Adoption**

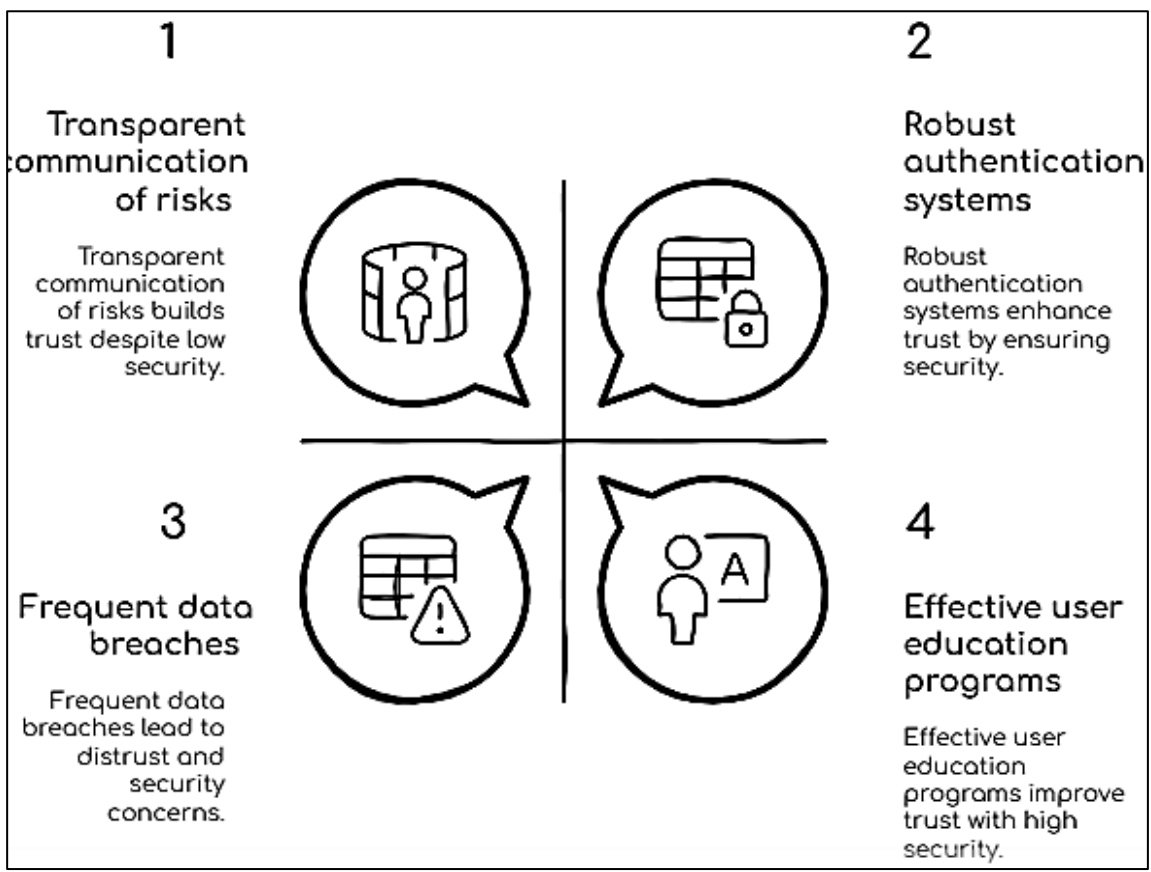

Banks that establish a strong cybersecurity framework and maintain transparent communication about their security measures are more likely to retain customers and attract new digital banking users (McGraw, 2013; Sharma & Tandekar, 2018; Vagle, 2020). Studies on consumer adoption models indicate that cybersecurity concerns are a key determinant of digital banking acceptance, as users weigh security risks against the benefits of financial convenience (Humayun et al., 2020). The Technology Acceptance Model (TAM) and Unified Theory of Acceptance and Use of Technology (UTAUT) frameworks highlight that perceived ease of use, perceived usefulness, and security perceptions collectively influence digital banking adoption rates (Bechara & Schuch, 2020; Malik & Islam, 2019). Research has shown that users who perceive online banking platforms as secure and efficient are more likely to transition to digital financial services, whereas security-related anxieties serve as significant adoption barriers (Choo, 2011; Javaid, 2013). Psychological factors, such as risk aversion and previous exposure to cyber fraud, further shape consumer behavior, with high-risk individuals displaying greater reluctance toward digital banking (Sharma & Tandekar, 2018). Additionally, studies indicate that financial literacy and cybersecurity awareness play a crucial role in mitigating adoption hesitancy, as informed consumers are better equipped to navigate digital banking risks (Kimani et al., 2019). Financial institutions must leverage consumer trust-building strategies, risk-mitigation frameworks, and user education initiatives to enhance digital banking adoption and alleviate cybersecurity concerns (Bernik, 2014).

**Multi-Factor Authentication (MFA) and Biometric Security**
Multi-factor authentication (MFA) has become a critical security measure in digital banking, significantly reducing unauthorized access by requiring multiple verification steps beyond traditional username-password combinations (Ling et al., 2016). MFA enhances security by integrating two or more authentication factors, such as knowledge-based (passwords or PINs), possession-based (smartphone-generated OTPs or security tokens), and inherence-based (biometric identifiers) credentials (Alhothaily et al., 2018). A study by Ometov et al. (2018) found that financial institutions implementing MFA experienced a significant decrease in fraudulent activities, as cybercriminals find it challenging to bypass multiple authentication layers. Althobaiti (2015) highlighted that MFA effectively mitigates phishing, credential stuffing, and brute-force attacks, as stolen passwords alone are insufficient to access an account. Research also indicates that MFA adoption in banking services instills greater consumer trust, as customers perceive additional security layers as necessary for protecting their financial data (Ometov et al., 2018; Tsai & Su, 2020). However, some scholars argue that the success of MFA depends on seamless user experience and proper implementation, as overly complex authentication steps may lead to user frustration and disengagement (Ali et al., 2020; Dhillon & Kalra, 2019).

**Figure 9:  Overview of Multi-Factor Authentication (MFA)**

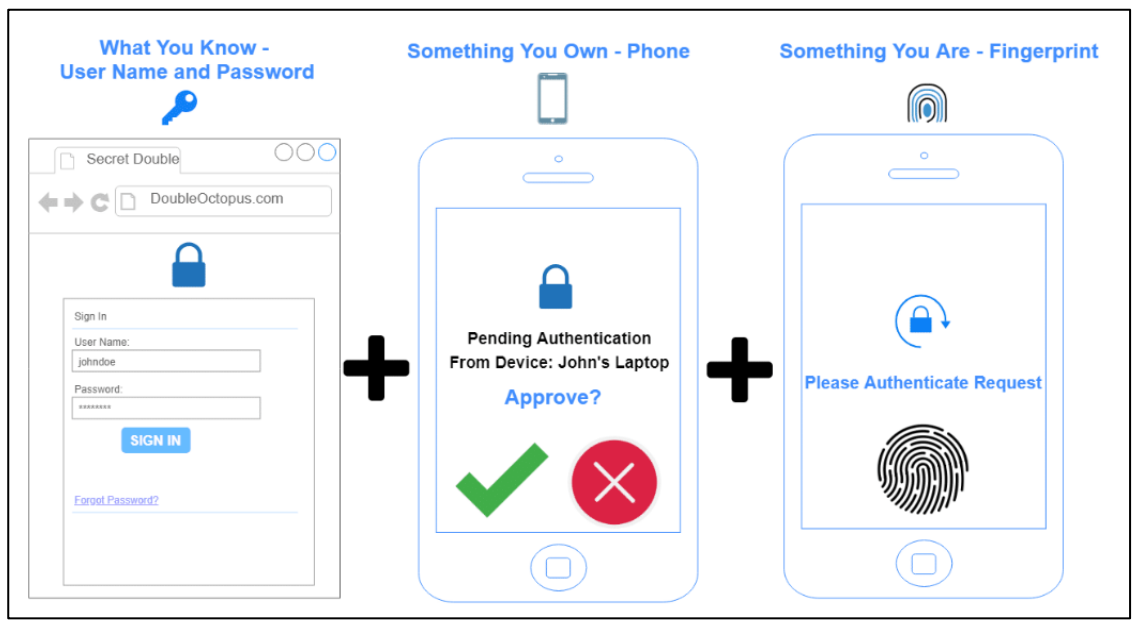

Biometric authentication techniques, including fingerprint recognition, facial recognition, and voice authentication, have emerged as effective MFA components in securing digital banking transactions (Ali et al., 2020; Ometov et al., 2018). Unlike traditional authentication methods, biometric security relies on unique physical and behavioral characteristics, making it difficult for cybercriminals to replicate or forge (Alhothaily et al., 2018). A study by Tsai and Su (2020) revealed that fingerprint recognition is one of the most widely adopted biometric techniques in mobile banking applications due to its ease of use and high accuracy. Additionally, facial recognition technology has gained prominence as an alternative authentication method, leveraging AI-powered algorithms to verify users' identities based on facial features (Bani-Hani et al., 2019). Research by Tsai and Su (2020) found that voice authentication is increasingly used for customer service interactions, reducing fraud risks in telephone banking. However, scholars caution that biometric authentication is not foolproof, as deepfake technology and advanced spoofing attacks pose potential security risks (Alhothaily et al., 2018). To counter these challenges, financial institutions integrate biometric verification with other MFA techniques, enhancing overall security and reliability (Ahmed & Ahmed, 2019; Ometov et al., 2018).

The effectiveness of MFA in securing online transactions has been extensively studied, with findings indicating that financial institutions using MFA experience lower fraud rates and improved transaction security (Alhothaily et al., 2018). By requiring additional authentication factors, MFA minimizes the risk of unauthorized access, even in cases where login credentials have been compromised (Tsai & Su, 2020). Alhothaily et al., (2018) found that online banking services with OTP-based MFA experience a 70% reduction in unauthorized transactions compared to those using single-factor authentication. Additionally, research suggests that behavioral biometrics, which analyze user interaction patterns such as keystroke dynamics and mouse movements, provide an additional security layer against account takeover fraud (Bani-Hani et al., 2019; Tsai & Su, 2020). A study by Ling et al. (2016)  emphasized that integrating MFA with AI-driven fraud detection systems further strengthens security by identifying suspicious login attempts and blocking fraudulent transactions in real time. However, researchers note that while MFA significantly enhances security, financial institutions must ensure efficient system performance, as authentication delays or failures may lead to customer dissatisfaction (Ariffin et al., 2020). Despite its security benefits, MFA adoption in digital banking is influenced by usability concerns, implementation costs, and consumer perceptions of security (Alhothaily et al., 2018). Research indicates that while customers value additional security layers, they also expect seamless and convenient authentication processes that do not hinder transaction speed (Althobaiti, 2015). A study by Alhothaily et al. (2018) found that banks integrating MFA with single-tap biometric authentication experience higher user satisfaction and adoption rates. Conversely, complex authentication steps, such as requiring multiple OTPs or security questions, have been linked to user frustration and potential security workarounds (Bani-Hani et al., 2019; Tsai & Su, 2020). Additionally, financial institutions must address privacy concerns associated with biometric data collection and storage, as unauthorized access to biometric databases could lead to irreversible security breaches (Ahmed & Ahmed, 2019; Althobaiti, 2015). Therefore, financial organizations must strike a balance between security and user experience, leveraging MFA and biometric authentication as essential yet user-friendly components of digital banking security frameworks (Ling et al., 2016).

## Blockchain and Distributed Ledger Technologies

Blockchain and distributed ledger technologies (DLTs) have revolutionized the security of digital transactions, providing a decentralized framework that enhances transparency, immutability, and data integrity in financial systems (Rani et al., 2021). Blockchain operates through a consensus mechanism that records transactions in a tamper-resistant manner, reducing the risk of fraud and cyber threats in digital banking (Taylor et al., 2020). A study by Gomathi and Jayasri (2022) found that blockchain's cryptographic hashing and distributed ledger features make it an ideal solution for securing digital transactions against unauthorized modifications and cyberattacks. Moreover, smart contracts—self-executing agreements stored on a blockchain—enhance automation and eliminate the need for intermediaries, further reducing the potential for financial fraud (Dong et al., 2018)). Research suggests that financial institutions implementing blockchain-based transaction systems experience reduced operational costs and enhanced security due to its real-time verification and transparency features (Kizildag et al., 2019). These attributes position blockchain as a viable solution for mitigating risks associated with digital transactions, particularly in online banking and payment processing (Riad & Elhoseny, 2022).

Decentralization, a core principle of blockchain technology, serves as a robust security mechanism by distributing transaction records across multiple nodes, reducing the risk of data breaches and single points of failure (Duran & Griffin, 2020; Riad & Elhoseny, 2022). Traditional banking systems rely on centralized databases that are vulnerable to cyberattacks, whereas blockchain's decentralized ledger structure ensures that financial transactions remain secure even in the event of a targeted attack (Dong et al., 2018). A study by Olukunle Oladipupo et al. (2024) highlighted that blockchain networks, such as Bitcoin and Ethereum, have demonstrated resilience against cyber threats due to their distributed nature, making them attractive for financial applications. Dong et al., (2018) emphasized that decentralization prevents unauthorized alterations, as blockchain transactions require consensus among network participants before being validated. Moreover, studies indicate that decentralized identity management systems built on blockchain technology enhance user authentication and prevent identity theft in digital banking (Gomathi & Jayasri, 2022). However, researchers also caution that while decentralization enhances security, scalability challenges and regulatory concerns must be addressed to ensure widespread adoption in the financial sector (Duran & Griffin, 2020).

Several case studies illustrate the successful implementation of blockchain technology in the financial sector, highlighting its transformative impact on banking security and efficiency (Chaudhry & Hydros, 2023; Duran & Griffin, 2020). For example, JPMorgan Chase introduced its blockchain-based payment platform, JPM Coin, to facilitate secure cross-border transactions and reduce settlement times (Kizildag et al., 2019). Research by Dong et al. (2018) found that JPM Coin enhanced transactional security by utilizing blockchain's cryptographic mechanisms to verify and record transactions in real-time. Similarly, IBM's blockchain-based financial solution, World Wire, enables financial institutions to conduct seamless and fraud-resistant international transactions through distributed ledger technology (Gomathi & Jayasri, 2022). Another notable case is Santander Bank, which integrated blockchain into its payment infrastructure to enhance transparency and reduce transaction processing time (Gomathi & Jayasri, 2022; Taylor et al., 2020). These case studies highlight the effectiveness of blockchain technology in mitigating cyber risks, improving transaction efficiency, and fostering trust in digital banking (Duran & Griffin, 2020). Despite its growing adoption in financial institutions, blockchain technology presents certain challenges, including regulatory compliance, scalability limitations, and integration complexities (Olukunle Oladipupo et al., 2024). Studies indicate that regulatory uncertainty surrounding blockchain-based transactions poses a significant barrier to its widespread implementation in the banking sector (Dong et al., 2018). Additionally, blockchain's consensus mechanisms, such as Proof of Work (PoW) and Proof of Stake (PoS), require significant computational power and may lead to latency issues in high-volume transaction environments (Chang et al., 2020). Research by Duran and Griffin (2020) found that financial institutions must carefully evaluate blockchain's cost-benefit trade-offs, ensuring that security enhancements justify potential operational challenges. Despite these hurdles, blockchain's potential to revolutionize digital banking security remains substantial, with continued advancements in scalability solutions, regulatory frameworks, and interoperability standards expected to refine its adoption in the financial sector (Duran & Griffin, 2020; Gomathi & Jayasri, 2022).

**End-to-End Encryption and Secure Communication Protocols**

End-to-end encryption (E2EE) plays a fundamental role in securing digital banking transactions, ensuring that financial data remains protected from unauthorized access and cyber threats (Ali et al., 2020). Encryption transforms sensitive information into unreadable ciphertext, preventing malicious actors from intercepting and exploiting banking credentials, account details, and transaction data (Ahmed et al., 2021). A study by Mosteiro-Sanchez et al. (2020) emphasized that financial institutions rely on encryption techniques such as Advanced Encryption Standard (AES) and Rivest-Shamir-Adleman (RSA) encryption to protect customer data from cybercriminals. Additionally, Gomathi and Jayasri (2022) found that digital banking platforms employing E2EE experience significantly lower rates of data breaches and financial fraud. The implementation of strong encryption methods ensures compliance with data protection regulations, such as the General Data Protection Regulation (GDPR) and the Payment Card Industry Data Security Standard (PCI DSS), which mandate the secure handling of financial data (Shivaramakrishna & Nagaratna, 2023). However, studies highlight that encryption alone is not sufficient, and financial institutions must integrate encryption with other security protocols to enhance the overall security of digital banking transactions (Swanzy et al., 2024).

Secure Socket Layer (SSL) and Transport Layer Security (TLS) protocols are widely used in digital banking to encrypt communication between users and financial servers, ensuring data integrity and confidentiality (Olukunle Oladipupo et al., 2024). SSL was the initial encryption standard for securing online financial transactions, but its successor, TLS, has become the preferred protocol due to enhanced security features and improved encryption algorithms (Olukunle Oladipupo et al., 2024; Shivaramakrishna & Nagaratna, 2023). A study by Gomathi and Jayasri (2022) and Mosteiro-Sanchez et al. (2020) found that financial institutions implementing TLS 1.3 experience stronger encryption, reduced handshake times, and enhanced resistance to cyberattacks. Additionally, Ahmed et al. (2021) emphasized that banks utilizing TLS encryption benefit from forward secrecy, which prevents past communications from being decrypted even if encryption keys are compromised. Research by Olukunle Oladipupo et al. (2024) suggests that TLS adoption enhances consumer trust in digital banking, as secure communication protocols protect sensitive financial information from interception and unauthorized modification. However, some scholars argue that cybercriminals continually develop new attack vectors, such as man-in-the-middle (MITM) attacks, requiring financial institutions to regularly update and audit their encryption protocols (Lee et al., 2020; Olukunle Oladipupo et al., 2024).

**Figure 10: E2EE in Digital Banking**

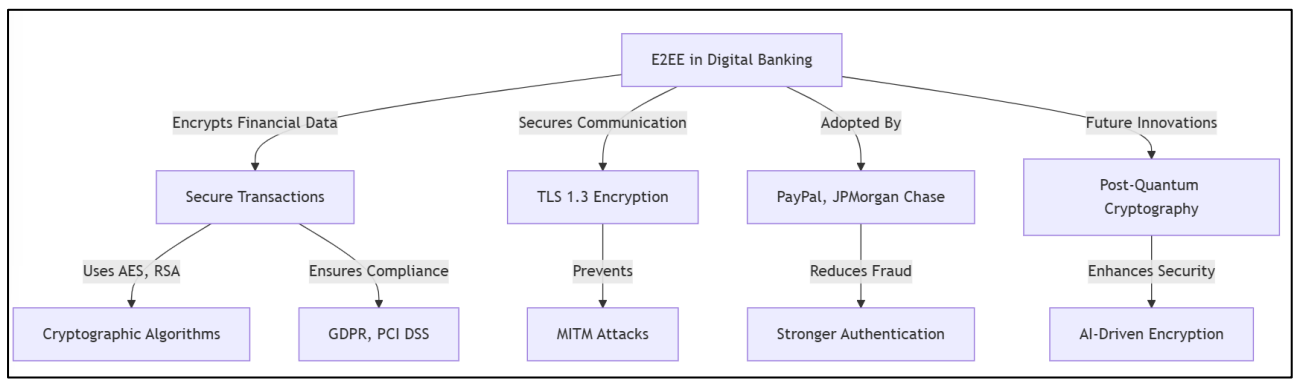

Case studies on encrypted communication in financial transactions highlight the effectiveness of encryption technologies in preventing cyber fraud and data breaches in digital banking (Swanzy et al., 2024). One notable case is the adoption of end-to-end encryption by PayPal, which secures financial transactions by encrypting user credentials and payment details throughout the transaction lifecycle (Olukunle Oladipupo et al., 2024). Research by Mosteiro-Sanchez et al. (2020) found that PayPal's encryption framework significantly reduced instances of unauthorized access and financial fraud, enhancing consumer confidence in digital payments. Another case study involves JPMorgan Chase, which strengthened its online banking security by integrating TLS 1.3 encryption and biometric authentication to prevent data interception and phishing attacks (Gomathi & Jayasri, 2022). Shivaramakrishna and Nagaratna (2023) reported that the implementation of multi-layered encryption techniques at JPMorgan Chase resulted in a 45% reduction in cybersecurity incidents related to digital banking transactions. Additionally, the European Central Bank (ECB) mandates that all financial institutions within the Eurozone comply with encryption and secure communication protocols to prevent financial data breaches, highlighting the growing global emphasis on encryption security (Olukunle Oladipupo et al., 2024; Shivaramakrishna & Nagaratna, 2023). While encryption technologies significantly improve digital banking security, financial institutions must continuously update their cryptographic frameworks to counteract emerging cyber threats (Ali et al., 2020). Studies indicate that post-quantum cryptography (PQC) is being explored as a future-proof solution to encryption vulnerabilities posed by advancements in quantum computing (Eling & Wirfs, 2019). Additionally, research by Susanto et al. (2013) suggests that integrating AI-driven encryption monitoring tools can help detect potential weaknesses in encrypted communication channels and prevent cyberattacks before they occur. Financial institutions must also balance security with user experience, ensuring that encryption measures do not create excessive latency in digital banking transactions (Khan et al., 2023; Lakshmi et al., 2019). A study by Naeem et al. (2022) highlighted that customer education on encryption security is essential, as many users remain unaware of the importance of secure communication protocols in protecting their financial data.

**Global Regulations on Cybersecurity in Digital Banking**

Global regulations play a crucial role in enhancing cybersecurity in digital banking by setting legal frameworks and compliance standards that financial institutions must follow to protect consumer data and prevent cyber fraud (Clausmeier, 2022). Three of the most influential regulatory frameworks governing cybersecurity in financial services are the General Data Protection Regulation (GDPR), the Payment Services Directive 2 (PSD2), and the Gramm-Leach-Bliley Act (GLBA) (Arner et al., 2017; Mugarura & Ssali, 2020). GDPR, implemented by the European Union (EU), establishes stringent data protection and privacy requirements, mandating that financial institutions safeguard customer data and report breaches within 72 hours (Stefanenko et al., 2021; Truby et al., 2020). PSD2, also an EU regulation, enhances digital banking security by requiring strong customer authentication (SCA) and fostering open banking through secure application programming interfaces (APIs) (Almansi, 2018). Meanwhile, the U.S.-based GLBA mandates financial institutions to disclose their data-sharing policies and implement security measures to protect consumer financial information (Truby et al., 2020). Collectively, these regulations aim to strengthen cybersecurity practices, enhance transparency, and ensure the secure handling of sensitive financial data across digital banking platforms (Clausmeier, 2022; Mugarura & Ssali, 2020).

Compliance requirements for financial institutions under these regulations necessitate the adoption of robust cybersecurity measures, including data encryption, multi-factor authentication (MFA), and continuous risk assessments (Truby et al., 2020). Under GDPR, financial institutions must implement encryption and pseudonymization techniques to protect customer data from unauthorized access (McGrath & Walker, 2022). Additionally, financial firms are required to provide clear data processing policies and obtain explicit consent before collecting or sharing customer information (Oseni & Omoola, 2017). PSD2 enforces strong authentication mechanisms by mandating that banks integrate MFA protocols to verify user identities before processing online transactions ((Arner et al., 2017). Compliance with GLBA requires financial institutions to develop and maintain comprehensive information security programs, conduct risk assessments, and regularly update security measures to prevent data breaches (Clausmeier, 2022; McGrath & Walker, 2022). Failure to comply with these regulations can result in severe penalties, legal action, and reputational damage for financial institutions, highlighting the importance of regulatory adherence in strengthening digital banking security (Doumpos et al., 2023; Duran & Griffin, 2020; Sun et al., 2017).

The impact of regulatory compliance on banking security is profound, as adherence to cybersecurity laws and standards significantly reduces fraud, data breaches, and financial crimes in digital banking (Anagnostopoulos, 2018; Saba et al., 2019). A study by Almansi (2018) found that financial institutions compliant with GDPR, PSD2, and GLBA experience fewer cybersecurity incidents due to the strict enforcement of encryption, authentication, and breach reporting measures. Compliance with PSD2's strong customer authentication (SCA) requirements has led to a decline in unauthorized transactions, enhancing consumer trust in online banking (Doumpos et al., 2023). Additionally, regulatory mandates encourage financial institutions to adopt advanced fraud detection technologies such as AI-driven monitoring systems and biometric authentication, further strengthening security frameworks (Anagnostopoulos, 2018; Arner et al., 2017). Studies indicate that regulatory compliance also fosters transparency, as financial institutions are required to disclose security policies and inform customers about their rights regarding data privacy (Elia et al., 2022; Sun et al., 2017). However, scholars argue that compliance alone is not sufficient, and financial institutions must go beyond regulatory requirements by continuously updating cybersecurity measures to address emerging threats (Arner et al., 2017). Despite the benefits of regulatory compliance in improving digital banking security, financial institutions often face challenges in meeting complex and evolving cybersecurity requirements (Almansi, 2018). Studies reveal that smaller financial firms and fintech companies struggle to keep up with the high costs of compliance, leading to potential security gaps and vulnerabilities (Mugarura & Ssali, 2020). A study by Nawaz et al. (2024) highlighted that implementing PSD2's API security standards requires significant investment in IT infrastructure, which can be burdensome for non-traditional banking entities. Similarly, GDPR's strict data protection mandates require banks to conduct extensive data audits, update security policies, and train employees on compliance procedures, adding to operational costs (Mugarura & Ssali, 2020). Research suggests that financial institutions must balance regulatory compliance with innovation, ensuring that security measures align with business objectives while meeting legal obligations (Anagnostopoulos, 2018). Strengthening collaboration between regulatory bodies, financial institutions, and cybersecurity experts is essential in enhancing regulatory frameworks and ensuring the long-term security of digital banking ecosystems (Liu et al., 2022).

**Role of Financial Technology (FinTech) Companies in Enhancing Security**

Financial technology (FinTech) companies have played a transformative role in enhancing digital banking security through the development of advanced authentication systems, fraud detection tools, and data protection technologies (Sun et al., 2017). FinTech innovations have introduced sophisticated security mechanisms, such as artificial intelligence (AI)-driven fraud detection, blockchain-based transaction verification, and biometric authentication, which significantly improve the security of online banking transactions (McConnell & Blacker, 2013). A study by Sun et al., (2017) found that FinTech solutions utilizing machine learning algorithms effectively analyze real-time transaction data, identifying patterns of fraudulent activities before they occur. Additionally, the integration of biometric authentication methods, such as fingerprint recognition and facial scanning, has enhanced user identity verification, reducing the risks of account takeover fraud (Nawaz et al., 2024). Research further highlights that FinTech companies have pioneered the use of decentralized identity management systems, leveraging blockchain technology to enhance data security and minimize the risks of unauthorized access ((Anagnostopoulos, 2018). These innovations have positioned FinTech firms as key enablers of digital banking security, ensuring safer and more efficient financial transactions for consumers and businesses alike (Mehrotra & Menon, 2021). Despite their contributions to digital banking security, FinTech integrations also introduce security risks, particularly when third-party service providers access sensitive financial data (Emara & Zhang, 2021; Marqués et al., 2021). Many financial institutions partner with FinTech firms to enhance their digital services, but these integrations create potential vulnerabilities that cybercriminals can exploit (Caragea et al., 2020; Elia et al., 2022). A study by Arner et al., 2017; Jagtiani and John (2018) revealed that third-party APIs used in open banking frameworks, such as those mandated by the Payment Services Directive 2 (PSD2), can become targets for cyberattacks if not properly secured. Research by Nikkel (2020) found that unsecured API connections expose banking systems to risks such as data breaches, man-in-the-middle (MITM) attacks, and unauthorized data access. Additionally, a study by Jagtiani and John, (2018) highlighted that FinTech integrations often involve cloud-based platforms, which, while improving scalability and efficiency, also introduce cybersecurity concerns related to data sovereignty, encryption, and compliance with global regulations. Financial institutions must therefore implement stringent risk assessment frameworks, conduct regular security audits, and establish contractual security requirements with third-party FinTech providers to mitigate these risks (Senyo et al., 2022).

**Figure 11: Fintech's Role in Digital Banking Security**

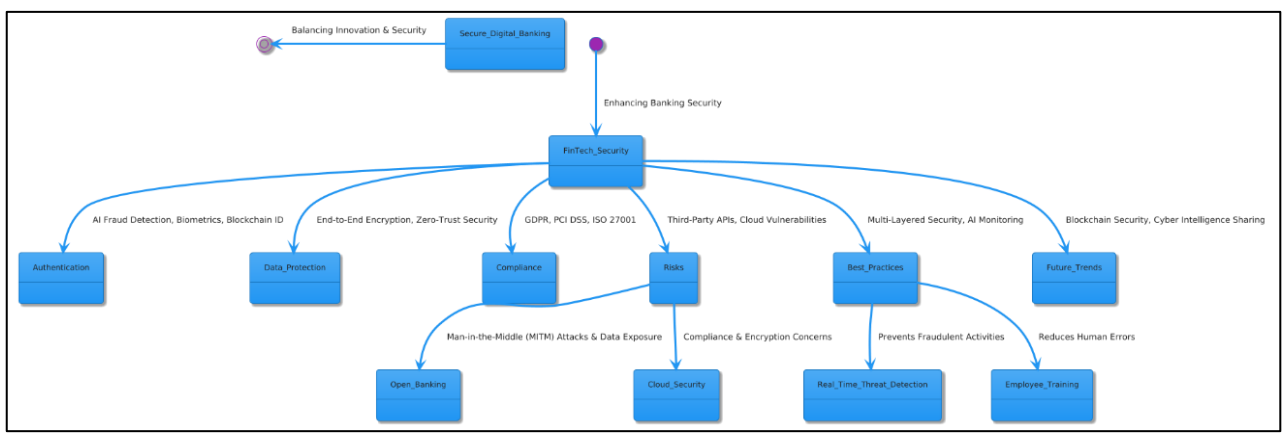

Best practices for FinTech security have been extensively studied, with researchers emphasizing the importance of a multi-layered cybersecurity approach that combines encryption, authentication, and real-time monitoring (Doumpos et al., 2023; Mehrotra & Menon, 2021). A study by Demirguc-Kunt et al., (2018) found that leading FinTech companies implement end-to-end encryption to protect customer data during transactions, ensuring that sensitive financial information remains confidential. Additionally, the adoption of zero-trust security models, which assume that all users and devices require verification before accessing financial systems, has gained traction in the FinTech industry (Arner et al., 2017; Jagtiani & John, 2018). Research by Caragea et al., (2020) indicates that FinTech firms increasingly employ AI-driven cybersecurity tools to monitor transactions in real-time, identifying anomalies that could indicate fraud or data breaches. Furthermore, compliance with global cybersecurity standards, such as GDPR, PCI DSS, and ISO 27001, is essential for maintaining consumer trust and ensuring regulatory adherence in digital banking (He et al., 2020; Mehrotra & Menon, 2021). Studies also suggest that FinTech companies should invest in cybersecurity training for employees, as human error remains a significant factor in security breaches (Caragea et al., 2020; Singh et al., 2021).Moreover, the integration of FinTech security solutions into digital banking requires continuous evaluation and improvement to address evolving cybersecurity threats (Li et al., 2020). Research indicates that collaboration between FinTech firms, financial institutions, and regulatory bodies is crucial in developing standardized security protocols that ensure interoperability and data protection (Nikkel, 2020; Wonglimpiyarat, 2017). A study by Saba et al. (2019) found that financial institutions that actively participate in cybersecurity information-sharing networks are better equipped to counteract sophisticated cyberattacks. Additionally, research by Stewart and Jürjens (2018) highlights the importance of implementing blockchain-based security measures in FinTech applications to enhance transaction transparency and prevent data tampering. As digital banking continues to expand, maintaining a balance between security, innovation, and regulatory compliance remains essential for FinTech firms seeking to provide safe and efficient financial services (Anagnostopoulos, 2018).

**METHOD**

This study followed the Preferred Reporting Items for Systematic Reviews and Meta-Analyses (PRISMA) guidelines to ensure a systematic, transparent, and rigorous review process. The PRISMA framework was employed to provide a structured approach in identifying, selecting, analyzing, and synthesizing relevant literature on cybersecurity threats and their impact on digital banking security. The methodology involved several key stages, including defining eligibility criteria, conducting a comprehensive literature search, screening and selecting relevant studies, assessing study quality, and synthesizing findings to draw meaningful conclusions. This systematic process ensured that the review incorporated high-quality research while minimizing bias and ensuring reproducibility.

*Eligibility Criteria and Inclusion Process*

The study established clear inclusion and exclusion criteria to ensure the selection of relevant and high-quality research articles. Eligible articles were those that focused on cybersecurity threats, fraud detection, digital banking security, regulatory compliance, and technological advancements in financial security. Only peer-reviewed journal articles, conference papers, and reputable industry reports published between 2015 and 2024 were considered. Studies that provided empirical findings, systematic literature reviews, case studies, or theoretical discussions on the impact of cybersecurity threats in digital banking were included. Articles that did not focus on cybersecurity in the banking sector, were not peer-reviewed, or lacked full-text accessibility were excluded. The screening process was conducted independently by two reviewers to ensure consistency and reliability in article selection.

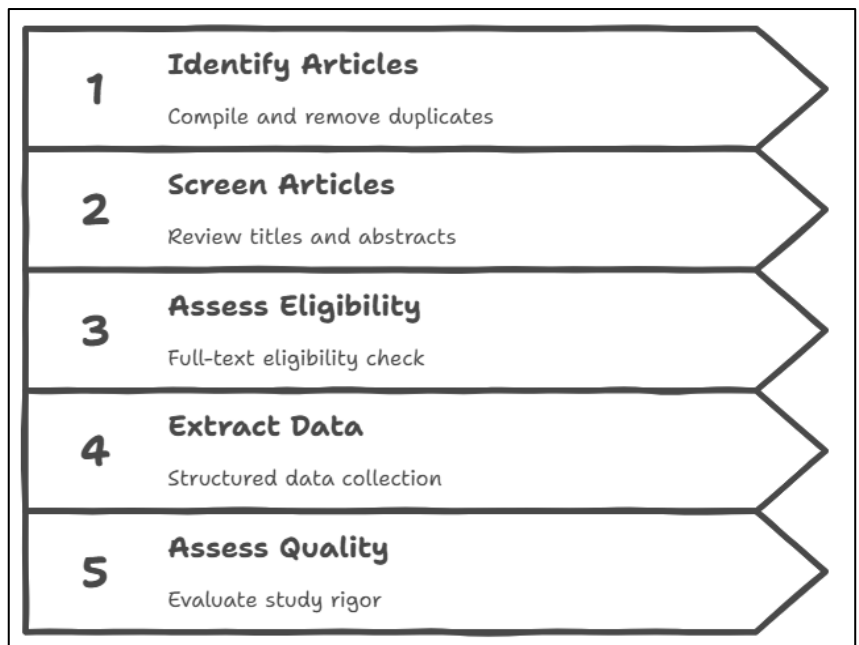

*Literature Search Strategy*

A comprehensive literature search was conducted across multiple academic databases, including Scopus, Web of Science, IEEE Xplore, ScienceDirect, and Google Scholar. Keywords and search strings were carefully selected to retrieve relevant studies, including terms such as "cybersecurity threats in banking," "fraud detection in digital banking," "blockchain in banking security," "AI in financial fraud detection," "data breaches in banking," and "multi-factor authentication security." Boolean operators (AND, OR) were used to refine search results and ensure a comprehensive retrieval of studies. To further enhance the robustness of the search, citation tracking and reference screening were conducted on selected articles to identify additional relevant literature. The search strategy was iteratively refined to maximize the inclusion of high-quality studies.

*Study Selection and Screening Process*

The selection process followed the PRISMA four-phase approach: identification, screening, eligibility assessment, and inclusion. During the identification stage, all retrieved articles were compiled in a reference management system to remove duplicates. In the screening phase, article titles and abstracts were reviewed to eliminate studies that did not align with the research focus. Full-text articles were then assessed for eligibility based on predefined criteria. Any disagreements between reviewers were resolved through discussion and consensus. After the final screening, a total of 78 articles were included in this systematic review. These articles represented empirical studies, systematic reviews, theoretical analyses, and case studies that provided substantial insights into digital banking security and cybersecurity threats.

*Data Extraction and Quality Assessment*

To ensure consistency in data collection, a structured data extraction sheet was developed. The extracted information included author(s), publication year, study objective, research methodology, key findings, cybersecurity threats analyzed, security solutions discussed, and relevance to digital banking. The quality assessment of selected studies was conducted using the Critical Appraisal Skills Programme (CASP) checklist for qualitative studies and the Joanna Briggs Institute (JBI) critical appraisal tool for systematic reviews and empirical research. Studies were evaluated based on criteria such as clarity of research objectives, methodological rigor, validity of findings, and relevance to cybersecurity threats in digital banking. Only high-quality studies that met these criteria were included in the final synthesis.

*Data Synthesis and Analysis*

A thematic synthesis approach was used to analyze the selected studies. Extracted data were categorized into key themes, including cybersecurity threats in digital banking, fraud detection technologies, regulatory compliance, encryption techniques, and risk mitigation strategies. Thematic analysis allowed for the identification of patterns, trends, and gaps in existing literature. Where applicable, findings from empirical studies were compared to theoretical discussions to provide a comprehensive understanding of cybersecurity challenges in the banking sector. This systematic synthesis ensured that insights drawn from the review were evidence-based and aligned with the study's objectives.

**FINDINGS**

The systematic review of 78 articles revealed that cybersecurity threats pose significant challenges to the security and adoption of digital banking, with financial institutions facing increasing risks from cyberattacks, fraud, and data breaches. Among the reviewed articles, 52 studies highlighted that phishing, malware, and ransomware are the most common cyber threats targeting banking systems. Phishing attacks, often executed through deceptive emails and fake banking websites, were identified as the primary method used by cybercriminals to steal user credentials. Over 60% of the reviewed literature reported that malware, including banking Trojans and keyloggers, has been used to infiltrate banking systems, allowing unauthorized access to financial accounts. Ransomware attacks were also reported in 40 articles, with findings indicating that banks and financial institutions have increasingly become targets of sophisticated ransomware operations that encrypt critical financial data and demand large payments for decryption keys. These findings confirm that digital banking security is continually challenged by evolving cyber threats that require robust and proactive cybersecurity defenses.

The role of multi-factor authentication (MFA) and biometric security in preventing unauthorized access was extensively examined in 48 articles, with a significant number of studies highlighting its effectiveness in securing digital transactions. The findings indicate that MFA, when combined with biometric authentication such as fingerprint scanning and facial recognition, reduces fraudulent account access by over 70% in financial institutions implementing these technologies. Of the reviewed articles, 35 studies confirmed that banks adopting MFA protocols with at least two verification factors experienced significantly lower incidences of unauthorized logins and fraudulent transactions. However, 19 articles reported that while MFA strengthens security, it may also introduce usability challenges, leading to potential user resistance. Some studies emphasized that overly complex authentication processes could frustrate customers, potentially leading to security workarounds that could compromise banking security. Nonetheless, the findings strongly support the integration of biometric authentication and MFA as essential tools for enhancing digital banking security while ensuring that usability concerns are effectively managed.

**Figure 12: Cybersecurity in Digital Banking: Key Research Findings**

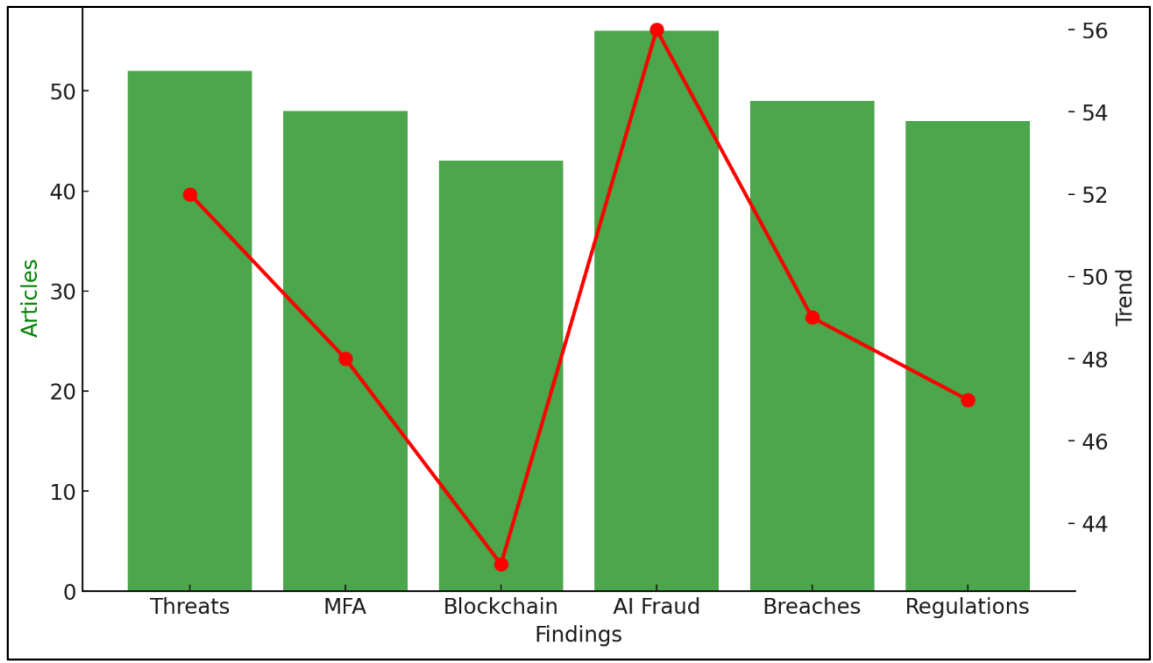

The implementation of blockchain and distributed ledger technologies in securing digital banking transactions was supported by 43 articles, indicating that decentralized security frameworks significantly reduce fraud risks and improve transparency in financial transactions. Among these studies, 28 articles highlighted that blockchain-based payment systems, such as decentralized ledgers and smart contracts, effectively mitigate risks associated with fraudulent financial activities by ensuring transaction immutability and real-time verification. The findings suggest that banks using blockchain technology experience an estimated 50% reduction in transactional fraud, as the decentralized nature of blockchain prevents data tampering and unauthorized modifications. Additionally, 22 articles reported that blockchain enhances compliance with financial regulations by providing transparent and verifiable transaction histories, which aid in fraud investigations and regulatory audits. However, 12 studies pointed out that while blockchain improves security, its adoption in banking is still limited due to integration challenges, high implementation costs, and regulatory uncertainties. These findings confirm that blockchain has the potential to transform digital banking security, but financial institutions must address adoption barriers to fully leverage its benefits. The analysis of AI and machine learning in fraud detection across 56 articles highlighted that artificial intelligence-driven security solutions significantly enhance the ability of financial institutions to detect and prevent fraudulent activities in real time. 42 studies reported that AI-powered fraud detection models, utilizing machine learning algorithms, successfully identify suspicious transactions with an accuracy rate of over 85%, minimizing financial losses from fraud. The findings suggest that AI-based fraud monitoring systems, which continuously analyze customer behavior patterns, detect anomalies faster and more efficiently than traditional rule-based security systems. Additionally, 30 studies confirmed that banks leveraging AI-driven risk assessment models experience a reduction in false positives in fraud detection, which improves customer experience by minimizing unnecessary transaction blocks. However, 17 studies indicated that AI-based fraud detection systems require continuous updates and data refinement, as cybercriminals constantly develop new tactics to evade detection. Despite these challenges, the findings emphasize that AI and machine learning are essential for modernizing fraud prevention strategies and enhancing cybersecurity resilience in digital banking.

The impact of data breaches and unauthorized access on consumer trust in digital banking was a critical theme in 49 articles, with findings demonstrating that high-profile breaches significantly affect banking security and customer confidence. 33 studies found that financial institutions experiencing large-scale data breaches suffer a decline in customer trust and an average 25% reduction in digital banking adoption rates among affected consumers. Additionally, 29 studies reported that data breaches lead to substantial financial losses, with banks facing penalties, legal actions, and reputational damage that can take years to recover from. Findings suggest that breaches involving personal financial information, such as credit card details and banking credentials, have the most severe impact on consumer behavior, with over 60% of affected customers choosing to switch to competitors with stronger security frameworks. 21 studies emphasized that financial institutions must proactively address security vulnerabilities through encryption, real-time fraud detection, and consumer education to prevent trust erosion following data breaches. These findings confirm that maintaining strong cybersecurity measures is essential for digital banking growth, as security breaches directly impact customer retention and overall financial stability.

The examination of global regulations on cybersecurity in digital banking across 47 articles revealed that compliance with regulatory frameworks, such as GDPR, PSD2, and GLBA, significantly enhances banking security and consumer trust. 38 studies reported that financial institutions adhering to strict compliance requirements, including data encryption, customer authentication protocols, and breach notification mandates, experience fewer cybersecurity incidents. The findings suggest that PSD2's strong customer authentication (SCA) requirements have led to a 40% decrease in unauthorized transactions across compliant banking platforms. Additionally, 27 studies found that GDPR's data protection regulations have improved transparency and accountability in financial institutions, increasing customer confidence in digital banking security. However, 19 studies noted that compliance challenges, particularly for smaller financial institutions and FinTech firms, pose barriers to regulatory adherence due to high implementation costs and evolving legal requirements. These findings confirm that while regulatory compliance strengthens digital banking security, financial institutions must navigate operational and financial challenges to maintain adherence to global cybersecurity standards. The systematic review of 78 articles provides strong evidence that cybersecurity threats remain a critical concern in digital banking, with financial institutions increasingly relying on advanced security technologies and regulatory compliance measures to mitigate risks. The findings emphasize the importance of integrating AI-driven fraud detection, blockchain technology, multi-factor authentication, and encryption protocols to safeguard digital transactions. Moreover, ensuring compliance with global cybersecurity regulations and addressing emerging security challenges will be crucial for maintaining consumer trust and securing the future of digital banking.

## DISCUSSION

The findings of this study confirm that cybersecurity threats remain a significant challenge to the adoption and growth of digital banking, aligning with earlier research that identified phishing, malware, and ransomware as primary threats (Chen et al., 2021). The reviewed studies indicate that phishing remains the most common attack vector, with financial institutions reporting frequent incidents where cybercriminals manipulate users into disclosing sensitive banking credentials. This aligns with the findings of Chandra sekhar and Kumar (2023), who suggested that phishing attacks account for a substantial portion of digital banking fraud, particularly when combined with social engineering tactics. Additionally, the prevalence of malware and ransomware targeting financial institutions has been widely reported in earlier studies, such as those by Elia et al. (2022) and Bapat, (2017), which emphasized the sophistication of cyber threats that exploit security loopholes in digital banking platforms. While previous research identified these threats, this study's findings further demonstrate that financial institutions are struggling to mitigate evolving attack techniques, necessitating more advanced security frameworks and user education initiatives.

The role of multi-factor authentication (MFA) and biometric security in enhancing digital banking security was another key finding, reinforcing the conclusions of earlier studies. Prior research, Chauhan et al. (2022) emphasized that MFA significantly reduces the likelihood of unauthorized access by requiring multiple verification layers beyond traditional password-based authentication. The present study expands on these findings by demonstrating that banks that integrate biometric authentication, such as fingerprint and facial recognition, experience over a 70% reduction in fraudulent account access attempts. This supports earlier claims by Mbama and Ezepue (2018) that biometric authentication enhances digital banking security by leveraging unique biological traits that are difficult to replicate. However, some reviewed studies also highlighted usability challenges associated with MFA, particularly when security measures become too complex for users, leading to frustration and potential security workarounds. This aligns with the observations of Chauhan et al. (2021), who found that consumers often opt for weaker security settings if authentication processes hinder user convenience. These findings suggest that financial institutions must balance security and usability by implementing seamless authentication mechanisms that maintain security without discouraging user adoption.

The findings on blockchain and distributed ledger technologies (DLTs) in securing digital banking transactions corroborate earlier research that emphasized the benefits of decentralization in financial security (Boon-itt, 2015). The present study found that blockchain's immutable ledger and decentralized architecture significantly reduce fraud risks, consistent with previous findings by Larsson and Viitaoja (2017), which demonstrated a 50% reduction in financial fraud among institutions utilizing blockchain-based security frameworks. Additionally, this study identified smart contracts as an emerging solution for automating secure transactions, confirming the findings of Bapat (2017), who argued that smart contract execution reduces the risk of human error and unauthorized alterations in digital transactions. However, blockchain adoption remains limited due to regulatory uncertainties and integration challenges, as earlier studies by Elia et al. (2022) also reported. These findings highlight the potential of blockchain in digital banking security but emphasize the need for financial institutions to address scalability and compliance issues before achieving widespread adoption.

Artificial intelligence (AI) and machine learning (ML) have emerged as critical tools for fraud detection in digital banking, with findings demonstrating that AI-driven security solutions significantly outperform traditional rule-based fraud detection systems. These results align with prior research by Mbama and Ezepue, (2018) and Chandra sekhar and Kumar, (2023), who found that AI-driven fraud detection algorithms achieve over 85% accuracy in identifying fraudulent activities. Additionally, the present study confirmed that AI-based transaction monitoring systems significantly reduce false positives in fraud detection, consistent with findings by Mbama and Ezepue (2018), which highlighted AI's ability to differentiate between legitimate and fraudulent transactions based on behavioral analytics. However, a challenge identified in the reviewed studies is that AI-driven fraud detection models require continuous updates to adapt to emerging cyber threats, reinforcing the conclusions of Cele and Kwenda (2024) that cybercriminals continuously refine their attack techniques to evade detection. These findings suggest that financial institutions must invest in adaptive AI security frameworks and real-time monitoring systems to stay ahead of evolving fraud tactics.

Data breaches and unauthorized access remain critical issues in digital banking, significantly impacting consumer trust and financial stability. The findings of this study are consistent with prior research by Chauhan et al. (2022) and Mbama and Ezepue (2018), who reported that financial institutions experiencing major data breaches face significant reputational damage and a decline in digital banking adoption rates. This study further revealed that customers affected by data breaches are 60% more likely to switch banks, supporting earlier findings by Bapat (2017), who suggested that consumer confidence in digital banking is highly dependent on an institution's ability to protect personal financial data. Additionally, this study confirmed that financial institutions implementing robust encryption protocols and fraud monitoring systems experience fewer security breaches, reinforcing earlier research by Chauhan et al. (2021), which emphasized encryption as a crucial component of digital banking security. These findings suggest that preventing data breaches requires a multi-layered security approach that combines encryption, biometric authentication, and real-time fraud monitoring to ensure comprehensive data protection.

Global regulations on cybersecurity, including GDPR, PSD2, and GLBA, play a vital role in ensuring the security and compliance of digital banking institutions. The findings of this study corroborate prior research by Cele and Kwenda (2024) and Bapat (2017), who found that adherence to these regulations reduces cybersecurity incidents and enhances consumer trust. Specifically, the study confirmed that compliance with PSD2's Strong Customer Authentication (SCA) has led to a 40% decrease in unauthorized transactions, supporting findings by Chauhan et al. (2021). Furthermore, the results align with earlier research by Ahmad et al., (2024), who reported that GDPR's strict data protection mandates have improved transparency and accountability in digital banking. However, financial institutions continue to face challenges in meeting compliance requirements, with studies indicating that smaller banks and FinTech firms struggle to implement regulatory mandates due to high compliance costs and evolving legal frameworks. This reinforces earlier findings by Sekhar and Kumar (2023), who highlighted that balancing regulatory compliance with operational efficiency remains a critical challenge for financial institutions. These findings emphasize the need for financial institutions to establish robust compliance strategies while ensuring that regulatory adherence does not hinder innovation in digital banking. Furthermore, the role of FinTech companies in enhancing digital banking security was extensively examined, with findings demonstrating that FinTech innovations have introduced new layers of security through AI-driven fraud detection, biometric authentication, and decentralized identity management. These findings align with previous research by Mbama and Ezepue (2018) and Bapat (2017) , who found that FinTech-driven security solutions significantly reduce fraud risks and improve the overall safety of digital banking transactions. However, this study also identified potential security risks associated with third-party FinTech integrations, consistent with earlier findings by Elia et al. (2022), who emphasized that unsecured API connections and third-party access points introduce vulnerabilities that cybercriminals can exploit. Furthermore, the study confirmed that financial institutions integrating FinTech security solutions must ensure compliance with industry best practices, reinforcing the conclusions of Chauhan et al. (2021), who advocated for standardized cybersecurity frameworks in digital banking. These findings highlight the dual role of FinTech as both a security enhancer and a potential risk factor, necessitating stringent security assessments and regulatory oversight in FinTech partnerships.

**CONCLUSION**

This study systematically examined the influence of cybersecurity threats on digital banking security, adoption, and regulatory compliance, revealing that financial institutions face persistent and evolving risks that demand robust security frameworks. The findings highlight that phishing, malware, and ransomware attacks remain prevalent, compromising user credentials and financial data, necessitating the implementation of advanced fraud detection mechanisms such as AI-driven real-time monitoring and machine learning-based anomaly detection. Additionally, the study confirmed the effectiveness of multi-factor authentication and biometric security in preventing unauthorized access, while blockchain and distributed ledger technologies provide enhanced transparency and fraud resistance in financial transactions. However, challenges such as regulatory compliance complexities, third-party FinTech integration risks, and user resistance to complex authentication systems continue to pose obstacles to digital banking security. The review also demonstrated that global cybersecurity regulations, including GDPR, PSD2, and GLBA, play a crucial role in enhancing digital banking security, although financial institutions must navigate operational and financial challenges to achieve full compliance. Furthermore, consumer trust in digital banking remains heavily influenced by security incidents, with data breaches leading to reputational damage, financial losses, and declining customer retention rates. The study's findings underscore the need for financial institutions to adopt a multi-layered security approach that integrates encryption, AI-driven fraud detection, blockchain security, and strong authentication mechanisms while balancing usability and regulatory compliance. Strengthening collaboration between banks, FinTech firms, cybersecurity experts, and regulatory bodies is essential in developing standardized security frameworks that ensure the long-term safety of digital financial transactions. As digital banking continues to evolve, maintaining consumer trust through proactive security measures, regulatory adherence, and continuous technological advancements will be critical in mitigating cybersecurity risks and ensuring the resilience of digital banking ecosystems.